\documentclass[11pt,preprint]{aastex}
\received{}
\accepted{30th May 2013}

\slugcomment{ApJ in press}
\shortauthors{Wilkes et al.}
\shorttitle{High-redshift 3CRR sources with \chandra}

\usepackage{graphics}
\usepackage{bm}
\usepackage{rotating}
\usepackage{times,mathptm,graphicx}
\usepackage{natbib}

\def\q24 {q$_{24}$}

\def\deg {$^{\rm o}$}

\def\xmm {{\it XMM-Newton}}

\def\nh {${\rm N_H}$}
\def\lax    {${_<\atop^{\sim}}$}
\def\gax    {${_>\atop^{\sim}}$}
\def\chandra {{\it Chandra}}

\def\tauSi {$\tau_{\rm 9.7 \mu m}$}
\def\rcd {$\rm R_{CD}$}
\def\5to8 {L$_{5\mu \rm m}$/L$_{8\mu \rm m}$}
\def\lxlr {L$_{\rm X}$/L$_{\rm R}$}
\def\lr {L$_{\rm R}(5~GHz)$}
\def\lx {L$_{\rm X}$}
\def\av {A$_{\rm V}$}
\def\spitz {{\it Spitzer}}

\begin{document}

\title{Revealing the heavily obscured AGN population
of High Redshift 3CRR Sources with \chandra\ X-ray Observations.}
\author{Belinda J. Wilkes$^1$, Joanna Kuraszkiewicz$^1$, 
Martin Haas$^2$, Peter Barthel$^3$, Christian Leipski$^4$, 
S.P.Willner$^1$, D.M.Worrall$^5$, Mark
Birkinshaw$^5$, Robert Antonucci$^6$, M.L.N. Ashby$^1$, 
Rolf Chini$^{2,7}$, G.G.Fazio$^1$, Charles Lawrence$^8$, Patrick
Ogle$^9$, Bernhard Schulz$^{10}$
}
\affil{1: Harvard-Smithsonian Center for Astrophysics, Cambridge, MA 02138}
\affil{2: Astronomisches Institut, Ruhr-University, Bochum, Germany}
\affil{3: Kapteyn Institute, University of Groningen, The Netherlands}
\affil{4: MPIA Heidelberg, Germany}
\affil{5: H.H. Wills Physics Laboratory, University of Bristol, UK}
\affil{6:Department of Physics, University of California, Santa
  Barbara, CA 93106}
\affil{7: Instituto de Astronom\'{i}a, Universidad Cat\'{o}lica del Norte,
Antofagasta, Chile}
\affil{8: JPL, Pasadena, CA 91109}
\affil{9: \spitz~Science Center, Caltech, Pasadena, CA 91125}
\affil{10: IPAC, Caltech, Pasadena, CA 91125}

\begin{abstract}
\chandra\ observations of a complete, flux-limited sample of
38 high-redshift (1$<$z$<$2), low-frequency selected (and so  
unbiased in orientation) 3CRR radio sources  are reported.
The sample includes 21 quasars (= broad line radio galaxies)
and 17 narrow line radio galaxies (NLRGs) 
with matched 178 MHz radio luminosity (log \lr$\sim 44-45$).
The quasars have high radio core-fraction,
high X-ray luminosities (log \lx$\sim 45-46$)
and soft X-ray hardness ratios (HR$\sim -0.5$) indicating
low obscuration. The NLRGs have lower core-fraction,
lower apparent X-ray luminosities (log \lx$\sim 43-45$) and mostly
hard X-ray hardness ratios (HR$> 0$) indicating obscuration
(\nh $\sim 10^{22-24}$ cm$^{-2}$). These properties and the correlation
between obscuration and radio core-fraction are consistent 
with orientation-dependent obscuration as in Unification models.
About half the NLRGs have soft X-ray hardness ratios and/or high $[$OIII]
emission line to X-ray luminosity ratio suggesting
obscuration by Compton thick (CT) 
material so that scattered nuclear or extended
X-ray emission dominates (as in NGC1068).
The ratios of unobscured to Compton-thin
($10^{22}<$ \nh(int) $< 1.5 \times 10^{24}$ cm$^{-2}$)
to CT (\nh(int)$> 1.5 \times 10^{24}$ cm$^{-2}$) is 2.5:1.4:1
in this high luminosity, radio-selected sample.
The obscured fraction is 0.5, higher than is typically reported for 
AGN at comparable luminosities from multi-wavelength surveys ($0.1-0.3$). 
Assuming random nuclear orientation, the unobscured half-opening angle 
of the disk/wind/torus structure is $\sim 60$\deg~and 
the obscuring material covers 30\deg~of which 
$\sim 12$\deg~is Compton thick.
The multi-wavelength properties reveal that many NLRGs have
intrinsic absorption $10-1000 \times$ higher
than indicated by their X-ray hardness ratios, and their true \lx~ values 
are $\sim$10--100$\times$ larger than  the
hardness-ratio absorption corrections would indicate. 

\end{abstract}

\keywords{Galaxies: quasars: general; X-rays: galaxies}

\section{Introduction}
\label{sec:intro}
The standard model for the nuclear regions of an active galaxy (AGN)
includes a super-massive black hole (SMBH)
surrounded by an accretion disk (AD) and corona  
producing strong, thermal optical-UV-soft X-ray and non-thermal X-ray 
emission. Gas and dust in the vicinity 
are heated by the nuclear emission 
producing the broad and narrow ultraviolet (UV), optical and infrared (IR)
emission lines, and near-IR hot dust emission characteristic of AGN. 
Radio-loud (RL) AGN also include relativistic jets of plasma streaming 
outwards from the nucleus along the accretion disk axis and emitting 
non-thermal synchrotron and associated inverse-Compton radiation.
The observed radio structures include jets, hot-spots, and lobes for which
the appearance (core-/lobe-dominated) 
is strongly related to the viewing-angle/orientation of the source to
our line-of-sight \citep{1989ApJ...336..606B}. AGN are
broadly classified into type 1 sources, with both broad and
narrow UV-IR emission lines, and type 2 sources with only narrow 
emission lines.
The detection of polarized broad lines in 
3C~234 \citep{1984ApJ...278..499A} and
NGC1068 \citep{1985ApJ...297..621A}
led to the general acceptance that some fraction of (narrow-lined) 
Seyfert 2s are absorbed, edge-on (broad-lined) Seyfert 1s, so that 
absorption and orientation are also factors for at least some 
radio-quiet AGN. The generally accepted
Unification model for AGN \citep{1989ApJ...336..606B,1985ApJ...297..621A}
interprets the observed range in emission line, radio structure, and other 
properties as being primarily due to the orientation of the source relative to
our line-of-sight.

As a result of the orientation-dependence of their observational 
characteristics, a critical problem in understanding AGN is to distinguish
observed differences due to orientation from  intrinsic physical 
differences. The AD and corona, possibly 
combined with a larger torus and/or wind
\citep{2000ApJ...545...63E,1994ApJ...434..446K},
provide obscuration which is anisotropic and strongly frequency-dependent
and results in complex, orientation-dependent selection effects 
for observations in most wavebands. This affects both source detection 
and classification. The orientation dependence of the 
observed Spectral Energy Distributions (SEDs) of AGN results in 
difffering levels of bias against most obscured sources in
traditional optical/ultraviolet/near-infrared/soft X-ray surveys.
Orientation unbiased surveys, which would properly test Unification
schemes, are difficult to come by. 

The advent of the Great Observatories has facilitated a number
of major multi-wavelength surveys 
(e.g.  SWIRE \citep{2003PASP..115..897L}, 
GOODS \citep{2004ApJ...600L..93G}, Bo\"{o}tes \citep{2007ApJ...671.1365H}, 
ChaMP \citep{2007ApJS..169..401K}, COSMOS \citep{2007ApJS..172....1S},
AEGIS \citep{2007ApJ...660L...1D,2004ApJS..154...48E}, 
CANDELS \citep{2011ApJS..197...35G}, HERMES \citep{2012MNRAS.424.1614O}) 
which, through the use of hard X-ray 
and/or mid-infrared (IR) selection, probe the AGN population, including 
obscured objects, more completely than traditional surveys
\citep{2000MNRAS.311...23B,2006ApJ...642..673D}.
Mid-IR selection requires secondary, usually X-ray selection
to distinguish AGN from the larger IR galaxy population
\citep{2012ApJ...748..142D,2006ApJ...642..673D,2006ApJ...642..126B},
but even then a bias against highly-obscured sources remains.
Although incomplete, surveys at near-IR wavelengths
(2MASS, \citet{2002ASPC..284..127C, 2005Natur.436..666M}) 
have revealed a population of red, moderately obscured AGN 
\citep{2002ApJ...564L..65W,2005ApJ...634..183W} of both types 
1 and 2 with space density comparable to normal type 1 AGN.
Current Cosmic X-ray Background (CXRB) 
models successfully include $\sim$equal populations of
unobscured and moderately obscured (log \nh(int) $\sim 21-23$)
AGN to model the emission up to $\sim 10$ keV \citep{2007A&A...463...79G}.
But a population of Compton thick (CT, \nh $> 1.5 \times 10^{24}$ cm$^{-2}$) 
AGN comparable to that of Compton thin
AGN is required to explain the higher energy ($\sim$30 keV) CXRB.
This CT population remains mostly undetected individually. 
They are difficult to
find, even at \chandra\ and \xmm\ X-ray energies (\lax 10 keV).
Direct light from NGC 1068, the ``Rosetta-stone" type 2 source,
is undetected by BeppoSAX {\it i.e.} to energies \gax 100 keV
\citep{1997A&A...325L..13M}. Estimates, which are 
based on the small numbers found and/or on X-ray stacking techniques
\citep{2012A&A...537A..16F,2009ApJ...693..447F,2007ApJ...670..156D,
2006ApJ...642..673D,2006A&A...451..457T,2006ApJ...636L..65B,
2006A&A...455..173P,2006A&A...446..459C,1999ApJ...522..157R},
cover a wide range 
($0.05 - >2 \times$ the rest of the AGN population).

Low-frequency radio-selection, which is based on extended, optically-thin
emission and so largely independent of orientation, provides the only
way to assemble a complete, randomly-oriented
sample of AGN. We are therefore carrying-out a multi-wavelength study of
a well-defined sample of low-frequency radio-selected (178 MHz), 
high-redshift (1$<$z$<$2), and thus high-luminosity (log \lr$\sim 44-45$), 
3CRR radio sources.
A major advantage provided by the radio data is an independent 
orientation indicator in the relative strength of (beamed) core and 
(isotropic) extended emission (core fraction 
\rcd~ = F$_{core}$/F$_{lobe}$(5~GHz),
\citet{1982MNRAS.200.1067O}). 
Models for the AGN nuclear obscuration range from geometrically thick, smooth
\citep{1988ApJ...329..702K} or clumpy \citep{2008ApJ...685..160N} tori 
to accretion disks with winds \citep{1994ApJ...434..446K,2000ApJ...545...63E}
and/or warps \citep{2010ApJ...714..561L}. 
The key input of the X-ray absorption column densities, 
IR-optical SEDs, and radio core fraction can help to discriminate between 
and/or constrain these models. 

One caveat to a radio-loud sample is that only $\sim 10$\% of AGN are 
radio-loud, and they may not accurately represent the majority AGN population,
{\it e.g.} radio emitting plasma may affect the opening angle of the
torus \citep{1995A&A...298..395F}, and generally the
X-ray emission includes an extra extended component related to the radio core
and jet.

\vskip -0.3in
\subsection{X-rays from radio-loud quasars.}
\label{sec:RLQ_X}
The X-ray emission from radio-quiet (RQ) AGN is well-known
to include multiple components \citep{1993ARA&A..31..717M}.
As well as the non-thermal, 
accretion-related power law which dominates the X-ray emission of 
luminous broad-lined AGN, contributions
from a soft excess, generally linked to the AD, reflected
emission from hot and/or cold material surrounding the nucleus, 
emission lines \citep{2003A&A...402..849O}, and/or
scattered nuclear light become significant in lower luminosity sources
and at high inclinations when the nuclear light is strongly obscured
\citep{1993ARA&A..31..717M}. For radio-loud AGN
(RLAGN) additional, non-thermal X-ray emission
is commonly associated with radio structures, lobes, and hot spots.
This can generally be resolved from the nuclear X-ray emission with the 
high spatial resolution of \chandra\ 
\citep{2012ApJ...745...84W,2009A&ARv..17....1W,2006ARA&A..44..463H}.
In the nucleus, the presence of an additional beamed, 
radio-jet-related component is demonstrated by the, on average,
$\sim 3 \times$ higher 
soft X-ray luminosity \citep{1981ApJ...245..357Z} and harder spectrum
\citep{1987ApJ...323..243W,1990ApJ...360..396W} of core-dominated (face-on), 
broad-lined RLAGN in comparison with RQAGN as observed with 
the {\it Einstein} Observatory. 

The strong correlation between core radio and X-ray luminosities
\citep{1999MNRAS.309..969H,1994ApJ...427..134W,1984ApJ...277..115F}
supports a Unification model in which beamed radio and X-ray 
emission originate at the base of the jet with the latter 
being related to the radio via synchrotron or synchrotron self-Compton 
processes. In lobe-dominated and edge-on sources, where a smaller 
beaming factor reduces the emission from this component, 
the X-ray emission lies above the X-ray/radio core correlation,
and the spectrum is softer, consistent with a significant contribution
from an accretion-related component as in RQAGN 
\citep{1999MNRAS.309..969H}. At low redshift (z$<$1) 
it has been possible to distinguish or place limits on the 
relative contributions from nuclear jet- and 
accretion-related X-ray components and confirm that the jet-related
component is more strongly related to the core radio emission
and absorption of the accretion-related component is related to 
source orientation
\citep{2009MNRAS.396.1929H,2006ApJ...642...96E,2006MNRAS.366..339B}.
The lower signal-to-noise (S/N) of the X-ray data for 
the higher redshift 3CRR sources presented here does not 
allow such separation.

\vskip -0.3in
\subsection{Orientation, Obscuration and Unification.}
\label{sec:unif}

While a level of Unification of luminous
quasars and radio galaxies is well-established \citep{1989ApJ...336..606B},
the variations in the relative numbers of quasars and radio galaxies as a
function of redshift and/or luminosity have called the simplest version of 
that scheme into question \citep{1993MNRAS.262L..27S,1982ApJ...256..410L}.
The ratio of obscured to all 
AGN (the ``obscured fraction'') remains a matter of debate as different
studies draw a variety of conclusions. At low redshift and luminosity,
optical surveys show that
type 2 AGN appear to be more numerous than type 1 by a factor of $\sim$a
few: obscured fractions of $\sim 0.65-0.75$ 
\citep{1995ApJ...454...95M,1992ApJ...393...90H,1982ApJ...256..410L}
are typical. Hard X-ray surveys, which are sensitive to 
gas rather than dust obscuration, find luminosity-dependent obscured
fractions of 0.2$-$0.8 at low-redshift (z \lax 0.1, INTEGRAL,
\citet{2012ApJ...757..181S}, {\it Swift}/BAT,
\citet{2011ApJ...728...58B}). 
High luminosity, radio-selected samples indicate 
an obscured fraction of $\sim$0.6
consistent with an unobscured half-opening angle of $\sim 53^o$
in Unification models \citep{2000MNRAS.316..449W} 
but with a luminosity dependence \citep{2005MNRAS.359.1345G} 
which can be explained by 
the ``receding torus model'' \citep{1995A&A...298..395F,1991MNRAS.252..586L}.
The 3CRR sample \citep{1983MNRAS.204..151L}
also shows a dependence on luminosity
with obscured fractions of 0.67 in the redshift range 0.5\lax z \lax 1 
\citep{1989ApJ...336..606B} and
0.5 in the current sample (1$\leq$ z $\leq$ 2).
Estimates based on the luminosity of the narrow, optical $[$OIII]$\lambda$5007 
emission line show a range of 0.4$-$0.9, also decreasing with luminosity
\citep{2005MNRAS.360..565S}.
X-ray surveys, again measuring the absorbing gas, 
generally conclude that the obscured fraction
decreases with luminosity and increases with redshift,
covering a range of $\sim 0.1-0.8$
\citep{2008A&A...490..905H,2006ApJ...652L..79T,2005ApJ...635..864L} although 
the redshift dependence may only be
present at high luminosities (\lx~$> 10^{44}$ erg s$^{-1}$,
\citet{2012A&A...546A..84I,2010AIPC.1248..359G}).
However \citet{2006MNRAS.372.1755D} find no relation between the obscured
fraction and either luminosity or redshift in their
analysis of deep \xmm\ observations of the \chandra\ Deep Field South (CDFS).
Estimates based on the IR emission imply a higher 
obscured fraction at high luminosity
($\sim 0.3-0.6$), a difference which may be due to missing 
highly-obscured sources in the X-ray surveys 
\citep{2008ApJ...675..960P,2008ApJ...679..140T}.

An alternate explanation for a luminosity dependence
of the obscured fraction 
is contamination by a second population of sources at low 
luminosity which are not standard, actively-accreting  
AGN \citep{2004MNRAS.349..503G,2000MNRAS.316..449W}.
There is a 
significant subset of low-luminosity NLRGs with low-ionisation 
emission lines (LERG, low-ionisation emission-line radio galaxy). 
Most FRI-type radio sources
\citep{1974MNRAS.167P..31F} and a subset of the lower-radio power 
(P$_{\rm 178MHz}$\,$<$\,10$^{\rm 26.5}$\,W Hz$^{-1}$ sr$^{-1}$)
FRII-type \citep{2002A&A...394..791C,2004MNRAS.349..503G}
are classified as LERGs. LERGS generally have weak, largely
unobscured X-ray emission 
\citep{2009MNRAS.396.1929H} and weak mid-IR emission 
(L(15$\mu$m) $< 8 \times 10^{43}$ erg 
s$^{-1}$, \citet{2006ApJ...647..161O}), both of which correlate
with AGN luminosity indicators such as $[$OIII]$\lambda$5007 
emission line luminosity (L$[$OIII])
\citep{2009ApJ...694..268D,1995ApJ...454...95M} and core 
radio strength. Thus there is
no evidence for a hidden, actively accreting nucleus, and
LERGs may be powered by a radiatively inefficient accretion flow 
\citep{2009MNRAS.396.1929H,2006ApJ...642...96E,2006ApJ...647..161O,2001A&A...379L...1G}.
In this case, the unresolved radio, IR, and X-ray cores would 
be purely jet- rather than accretion-related, e.g. as in 
M87 \citep{2004ApJ...602..116W,2000MNRAS.316..449W}, and LERGs would
not be part of the primary AGN population.
The obscured fraction in the 3CRR sample determined without the LERGs
is $\sim 0.5-0.6$ with little/no dependence on redshift or luminosity
\citep{2006ApJ...647..161O,1989ApJ...336..606B}. 
It is clear that
studies of the obscured fraction as a function of luminosity and z need to
take account of source classification.

\spitz~studies of luminous 3CRR sources, i.e. excluding 
LERGs, from $0.05<$z$<2.0$ show no luminosity dependence
of the obscured fraction and thus support simple Unification. 
At shorter wavelengths, our multi-wavelength study 
of the high-redshift (z$>$1) 3CRR radio sample has demonstrated 
a marked difference between the \spitz-observed IR SEDs of radio galaxies 
and quasars \citep{2008ApJ...688..122H,2010ApJ...717..766L,
2006ApJ...647..161O,2005ApJ...629...88S} which can be
explained by nuclear obscuration of a randomly-oriented sample
in a Unification model. Studies of lower-redshift 3CRs agree 
but also show evidence for a contribution
to the inclination dependence from beamed emission \citep{2007ApJ...660..117C}.
At longer wavelengths, 24$\mu$m and 70$\mu$m, emission is unrelated
to the source orientation \citep{2009ApJ...694..268D,2004A&A...424..531H}.
Herschel observations suggest
a significant contribution by star formation, expected to be independent
of obscuration, in the far-IR for a subset of the sources
\citep{2012ApJ...757L..26B}, supporting results by \citet{2007ApJ...661L..13T}.
The inner parts of the narrow emission line regions (NLR) may also be
obscured, weakening the $[$OIII]$\lambda$5007 emission line
\citep{2005A&A...442L..39H,1990Natur.343...43J} but not $[$OII]$\lambda$3727
\citep{1993Natur.362..326H}, and resulting in the
higher-ionization lines being 
visible only in the IR. However, for the highest luminosity radio sources
this seems not to be a large effect 
\citep{2004MNRAS.349..503G,1997MNRAS.286..241J}, perhaps due to the
more extended and so less obscured NLR in 
higher luminosity radio sources \citep{2000MNRAS.311...23B}.

\vskip -0.3in
\subsection{This paper}

The high-redshift 3CRR sample, which includes only powerful,
actively accreting AGN (no LERGs) with a limited range of both 
luminosity and redshift, is a particularly uniform
and well-suited sample with which to investigate the
relation of the full SED to orientation/obscuration and to study
the properties of the obscuring material. 
This paper describes the 3CRR 
sample (Section 2), presents our analysis of new and existing 
\chandra\ and XMM-Newton data
(Section 3), characterizes the X-ray properties and 
investigates their relation to radio and IR emission (Section 4), 
discusses the results 
in the context of Unification models (Section 5),
and summarizes the conclusions (Section 6).

Throughout the paper we assume a $\Lambda$CDM cosmology 
with H$_o$=71 km s$^{-1}$ Mpc$^{-1}$,
$\Omega_M=0.27$, and $\Omega_{\Lambda}=0.73$ \citep{2011ApJS..192...16L}.

\section{The Sample and Supporting Data}
\label{sec:sample}

The 3CRR catalog \citep{1983MNRAS.204..151L}
contains 180 radio galaxies and Quasi-stellar Radio Sources, 
quasars, up to redshift $z$\,=\,2.5 and is
100\% complete to a flux of 10 Jy at 178 MHz .
At these low frequencies all sources are dominated by emission
from the radio lobes resulting in little/no bias based on the orientation of 
the source. The 3CRR sample has been studied in detail over
many wavebands. The radio morphologies 
are well known and their radio sizes, lobe separations and jet prominence,
and core fractions at higher frequencies (5 GHz)
permit estimates of their radio axis orientation.
We have selected a complete sample of high-redshift (1\,$<$\,$z$\,$<$\,2) and
thus high luminosity, 3CRR
sources to ensure they are actively accreting, that none are LERGs
\citep{1979MNRAS.188..111H}, and
that the AGN dominates the bolometric luminosity of the source.
The complete sample of 38 3CRR high-z sources\footnote{which includes two 
4C sources found by \citet{1983MNRAS.204..151L}
to match the 3CRR selection criteria} (Table~\ref{tb:obs}) includes
21 lobe-dominated, steep spectrum quasars (a.k.a. broad-line radio galaxies, 
Quasi-Stellar Objects, QSOs) and 
17 narrow line radio galaxies (NLRGs), all of
Fanaroff-Riley type FR\,II with 
double lobes of P(178~MHz) $>$\,10$^{\rm 26.5}$\,W Hz$^{-1}$
generally extending far beyond the host galaxy. A subset of both types 
(6 quasars, 2 NLRGs) has steep radio spectra ($\alpha > 0.5$)
and compact ($<$10 kpc) structure (CSS, compact steep spectrum, 
\citet{1998PASP..110..493O,1985A&A...143..292F}).
There is at most one marginally 
core-dominated radio source, 3C~245,\footnote{A compact 
triple source with a steep
radio spectrum \citep{1984A&A...140..399B,1990A&A...228...17F} for which
a variable core may result in a high core-fraction.}
in this sample so that beamed emission is not generally dominant. 

Because of their brightness (F(178~MHz) $>$\,10\,Jy), the 
complete nature of the survey, the comprehensive multi-wavelength data
available, and their high luminosity, the high-redshift 3CRR sources
constitute an excellent sample with which to study orientation-based
effects. The mean 5\,GHz luminosity 
(log $\nu$ \lr~$\sim 44.5$ erg s$^{-1}$, Figure~\ref{fg:LrLx}(left))
is about five times higher than for the 3CRR sources at 0.5\,$<$\,$z$\,$<$\,1. 
The high redshift 
lowers the effects of X-ray absorption, which largely 
shifts out of the \chandra\ band unless the source is close to CT.
This strong negative k-correction means that
the X-ray flux of heavily absorbed AGN is not such a strong function of
redshift in this range \citep{1999MNRAS.309..862W}, 
and CT AGN are detectable.
\spitz~IRAC and MIPS photometry have been obtained to delineate the IR 
continuum for the full sample and IRS spectroscopy for those in the
range 1.0$<$z$<$1.4  
\citep{2008ApJ...688..122H,2010ApJ...717..766L}.

\section{X-ray Data and Analysis}
\label{sec:xrayanalysis}

Eleven sources in our high-z 3CRR sample, 2 NLRGs 
and 9 quasars, 
were previously observed by \chandra\ and 4 sources, 
3 NLRGs 
(1 in common with \chandra) and 1 quasar, 
with \xmm. New \chandra\ ACIS-S observations
were obtained for the remaining 24 quasars and NLRGs and for 3C~270.1
where the existing \chandra\ data were of poor quality.
The exposure times were set for a detection at expected levels for
NLRGs and quasars as a function of redshift.
Sub-arrays were used for the brightest quasars to avoid pileup issues.
The observations used in this paper, both new and archival, are listed in 
Table~\ref{tb:obs} along with known properties of the sources
and references to published analysis of existing 
\chandra\ and/or \xmm\ X-ray data.
All but one sources were detected making this the most complete 
X-ray-observed sample of AGN to date. There is a wide range of 
S/N from a
few counts for the faintest NLRG to $\sim 1000$ net counts for the brightest
quasars.

The \chandra~data were processed using the standard pipeline 
with calibration products appropriate for their observation dates. 
Archived \chandra~datasets 
observed in ACIS VFAINT mode were reprocessed to take advantage of improved 
calibration of the CTI correction and background cleaning. Counts 
were extracted from a 2.2$''$ radius circle (to enclose the full
point spread function) centered on the X-ray source
position for energy bands: broad (B, 0.3$-$8.0 keV), soft (S, 0.3$-$2.0
keV), and hard (H, 2.0$-$8.0 keV). 
Background counts were extracted in the same energy bands
from an annulus centered on the position of each 
source with inner and outer radii of 15$''$ and 35$''$
respectively, adjusted if necessary to exclude nearby sources.
Any sources remaining within the background annulus were
removed. The resulting net counts  
and hardness ratios\footnote{Hardness ratio based on the counts, 
HR=(H-S)/(H+S) with uncertainties 
determined using the Bayesian estimation (BEHR)
method, \citep{2006ApJ...652..610P}} 
are given in Table~\ref{tb:flux}.

In order to provide a uniform set of derived X-ray properties,
all \chandra-observed sources were run through an automated spectral 
analysis process using the Levenberg-Marquardt optimization method
in CIAO/Sherpa with the $\chi^2$ statistic including 
the Gehrels variance function, 
which allows for a Poisson distribution for low-count sources.
Two spectral fits were performed, the first fit assumed a power law 
spectrum with a canonical slope $\Gamma=1.9$ 
\citep{2007ApJ...665.1004J,1993ARA&A..31..717M}
and Galactic absorption as characterized by the equivalent hydrogen
column density (\nh(gal), Table~\ref{tb:obs}, \citet{1990ARA&A..28..215D}) 
the second added an intrinsic absorption component
at the redshift of the source (\nh(int)). The results were inspected
individually, and those for the best spectral fits
are listed in Table~\ref{tb:flux}, including
Galactic and intrinsic absorption-corrected fluxes and luminosities
in standard bands. When no significant \nh(int) was
detected a $3 \sigma$ upper limit is listed. 
Spectral fits for sources with low net counts ($<50$)
provided no useful constraint on \nh(int).
For sources where the data have sufficient 
net counts ($>700$), mostly quasars, the results of
a third spectral fit allowing the power law slope to be free are also listed 
in Table~\ref{tb:flux}.
Derived spectral parameters are consistent with those reported in 
published data except where noted in the table. 
Detected \nh(int), indicating absorption in excess of the Galactic 
column density, is most likely to be absorption
intrinsic to the quasar associated with 
the nucleus and/or the host galaxy. Although unlikely, a contribution
from absorption by intervening material/sources along the line-of-sight
cannot be ruled out.

There are 3 sources with only \xmm\ data: 3C~239/322/454.0.
The results of published 
spectral analysis were used to derive equivalent \chandra\ quantities 
for 3C~239/454.0 \citep{2008A&A...478..121S}. 
For 3C~322 the \xmm\ data showed no detection of the AGN
\citep{2004MNRAS.352..924B}. In order to 
determine an upper limit, the data were measured directly
(Tables~\ref{tb:obs},\ref{tb:flux}).


About half the sample show significant extended X-ray emission.
An example is 3C~270.1 \citep{2012ApJ...745...84W}
with X-ray emission related to the
radio structure, as is often observed 
\citep{2006ARA&A..44..463H,2009A&ARv..17....1W}, 
along with possible detection of
thermal emission from a surrounding cluster. 
The study of the extended X-ray emission will be covered in a later paper.
Extra-nuclear emission
originating close to the nucleus will generally not be resolved
at these high redshifts.
In cases with visible extent which may be contaminating the 
nuclear X-ray counts determined via the standard extraction region, an 
additional, smaller region was also used to better isolate the emission from 
the nucleus. When the results differ, a second set of counts and hardness 
ratios is reported in Table~\ref{tb:flux}, footnote 12. These counts were 
not used in our general analysis in order to ensure uniform measurements
across the full sample.

\section{Results}

\subsection{X-ray Luminosity}

\begin{figure}
\epsscale{1.0}
\plottwo{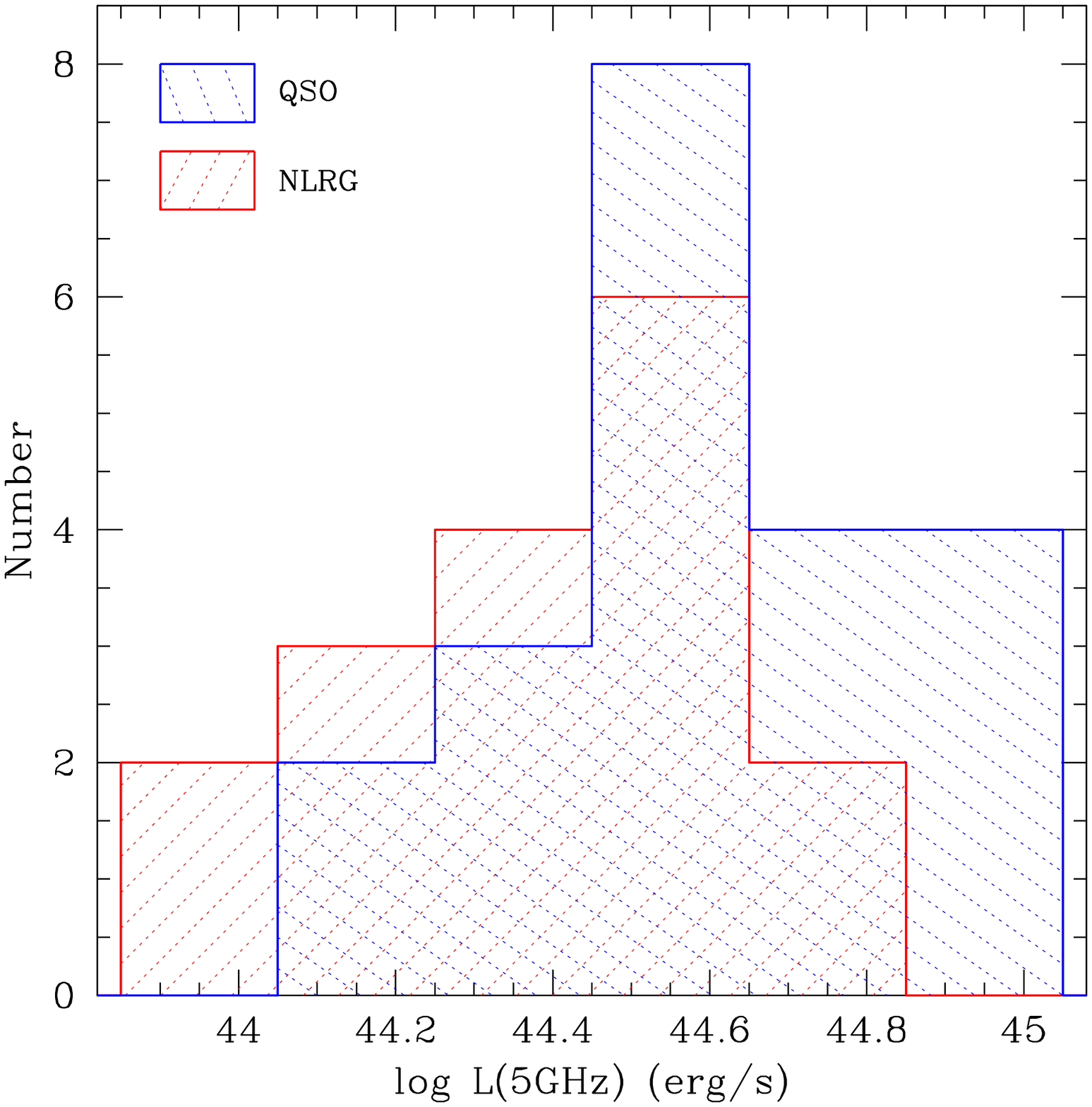}
{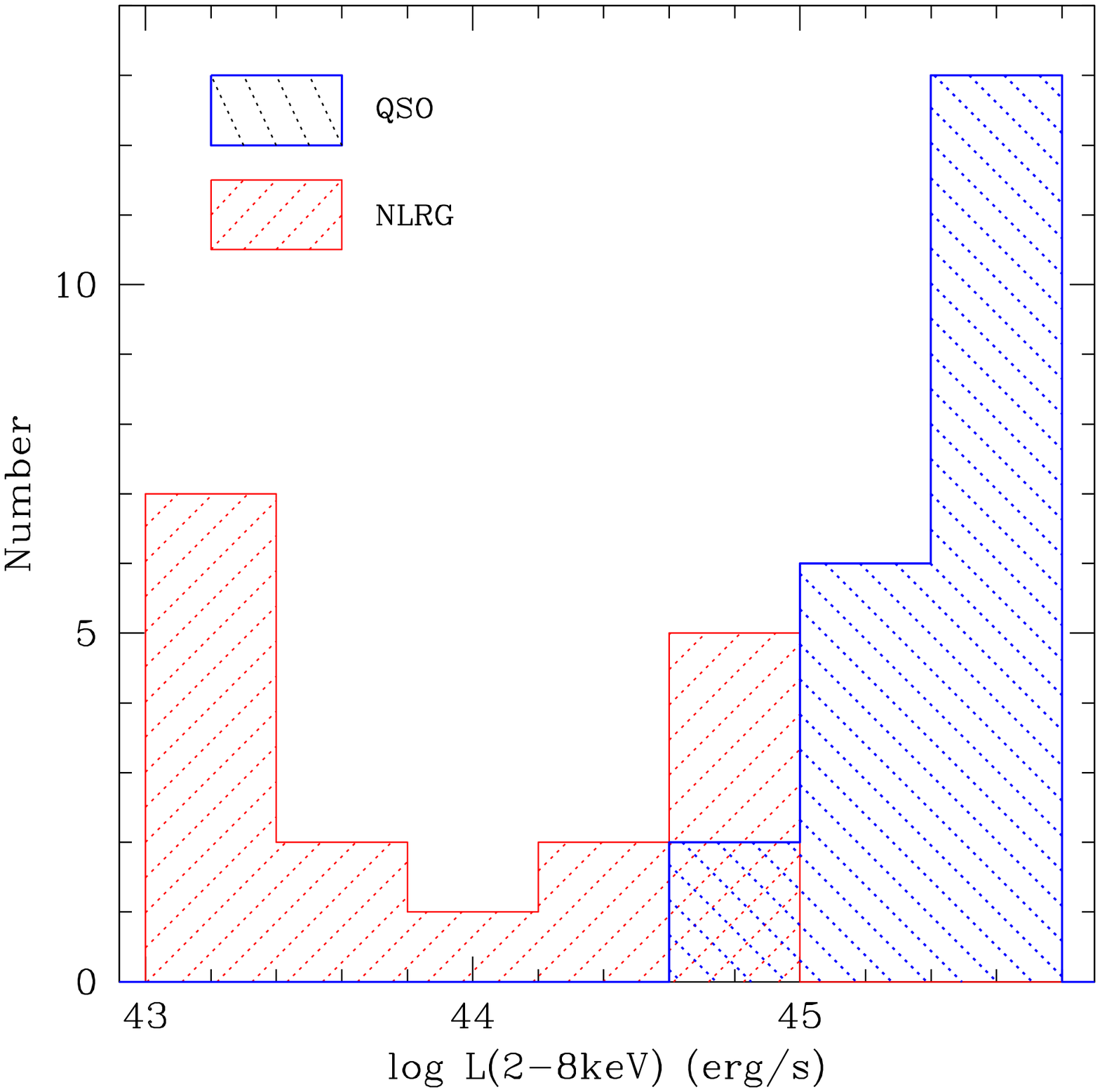}
\caption{A comparison of the distributions of total rest-frame 5 GHz 
radio (left) and hard-band, nuclear 
X-ray (right) luminosities, uncorrected for intrinsic absorption, 
for the quasars (QSOs, blue) and NLRGs (red).
In the radio the range is small (within 1 dex) with a small shift to 
higher luminosities for the quasars indicating a $\sim 30$\%
contribution from the beamed core at this relatively high frequency.
In the X-rays the full distribution covers
$\sim$2.5 dex and the quasars are easily distinguishable by their
brighter X-ray emission.
}
\label{fg:LrLx}
\end{figure}

Being low-frequency (178 MHz) radio-selected, for which the emission is
generally optically thin, the quasars and NLRGs are 
well-matched in this orientation-independent parameter.  
Figure~\ref{fg:LrLx}(left) shows the distributions of 
total radio luminosity at the higher frequency of 5 GHz, where beaming 
is more important. The overlap remains good, with a small shift 
towards brighter luminosities for the quasars.
The difference between the median luminosities of the 
NLRGs (log \lr~= 44.41), and the quasars (log \lr~= 44.59) indicates 
that beamed emission 
from the core contributes on average $\sim 30$\% of the radio luminosity 
in the lobe-dominated quasars and likely a similar fraction of the X-ray 
luminosity as well.

The distribution of
X-ray luminosities derived from the initial power law spectral
fits, with no \nh(int) included, is shown in Figure~\ref{fg:LrLx}(right). 
In contrast to the radio luminosity,
the X-ray luminosity distributions barely overlap, demonstrating the well
known difference between the observed X-ray emission from quasars and NLRGs 
with the quasars factors of $\sim 10 - 1000$ brighter
\citep{1999MNRAS.309..969H,1994ApJ...420L..17W}. 
In this sample, the ratio of the median \lx\ for quasars and NLRGs is 
$\sim 100$. Unification models interpret this difference as due to 
obscuration in the edge-on NLRGs.

\subsection{X-ray Hardness Ratio and Absorption}
\label{sec:xrayabs}

\begin{figure}
\epsscale{1.0}
\plotone{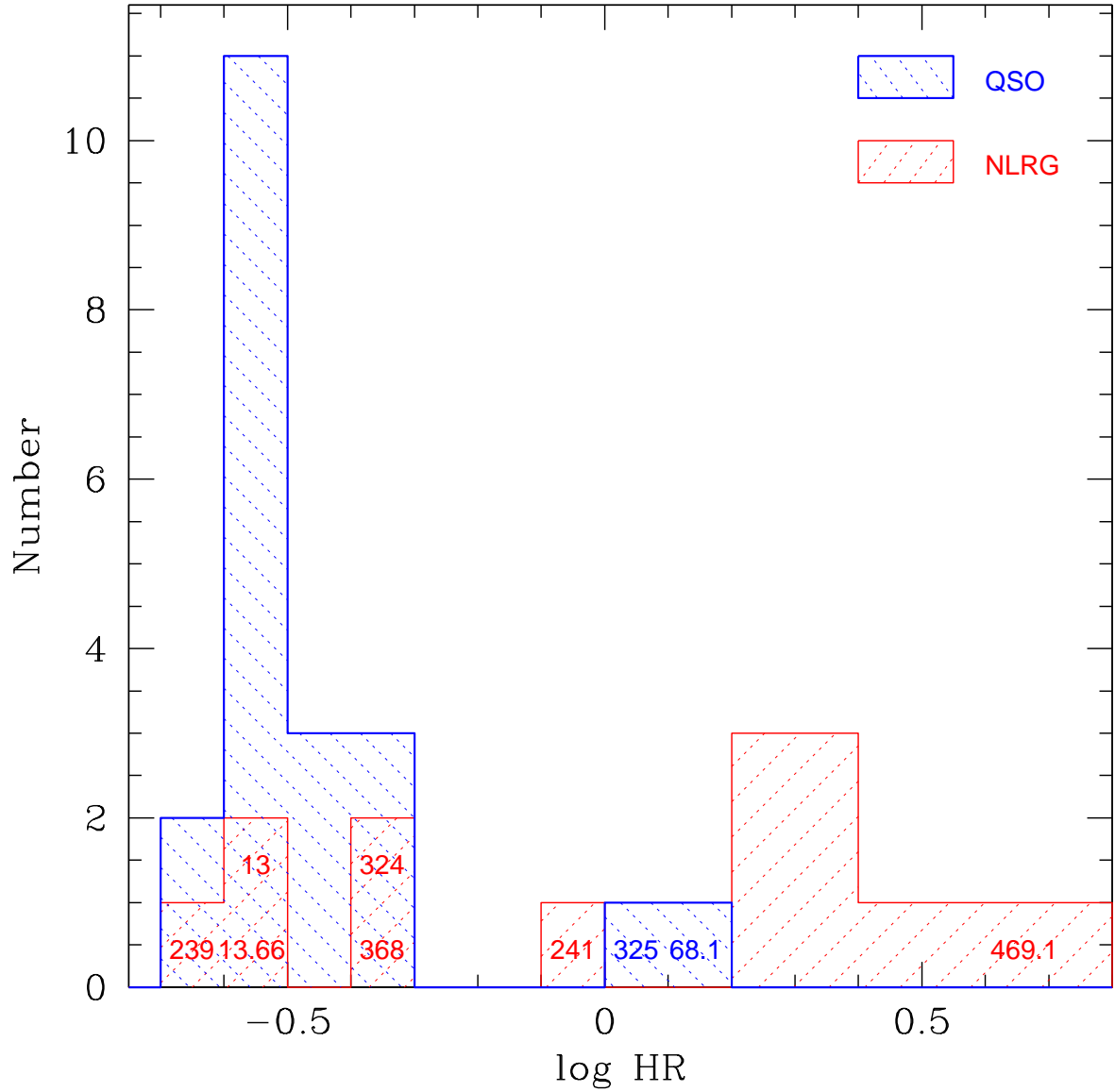}
\vskip -1.5in
\caption{Histograms of the X-ray hardness ratios 
for quasars (QSOs, blue) and NLRGs (red)
showing soft emission for all but two of the quasars and a wide range of
hardness ratio for the NLRGs.
}
\label{fg:HRdist}
\end{figure}

X-ray hardness ratio is an indicator of the intrinsic spectrum
which can be used over a wide range of S/N. 
Assuming the primary power law dominates and that its spectral index is similar
in all sources, the hardness ratio statistically 
indicates the amount of obscuration. 
Figure~\ref{fg:HRdist} shows a comparison of the distribution of X-ray
hardness ratios determined using our standard X-ray analysis 
(Table~\ref{tb:flux})
for the quasars and NLRGs in the 3CRR high-redshift sample. The
quasars, with only two exceptions, show soft spectra covering a narrow
range of hardness ratio ($\sim -0.5$), 
consistent with the average spectrum of a
quasar: $\Gamma \sim 1.9$ and little/no obscuration. 
The NLRGs, on the other hand, cover a wide range of hardness ratios
($-0.7 <$HR$< 0.7$). Five are consistent with the soft
spectra of quasars, but the majority  are significantly harder. The two
harder quasars noted above 3C~68.1/325,  along with NLRG 3C~241
lie between the quasars and NLRGs, having moderately hard
spectra consistent with obscuration by material with \nh(int) $\sim$
$10^{22-23}$~cm$^{-2}$.

The X-ray spectra of quasars are known to include several components,
limiting the accuracy with which any intrinsic 
obscuration can be determined in low S/N data (Section~\ref{sec:intro}).
For example, the presence of a soft excess or scattered
nuclear emission decreases the estimated obscuration if not accounted
in the fits. By contrast, 
the presence of a reflection component would harden the effective X-ray slope 
at high energies, resulting in an over-estimate of the 
obscuration in our single power law fits to individual sources. 
Thus both hardness ratios and single power law spectral fits can be
misleading in individual cases \citep{2005ApJ...634..183W,2005MNRAS.362..784P}.
Higher S/N data than are available for the
NLRGs in this sample are required to reliably
de-convolve any multiple spectral and/or spatial components. 

\begin{figure}
\epsscale{0.9}
\plotone{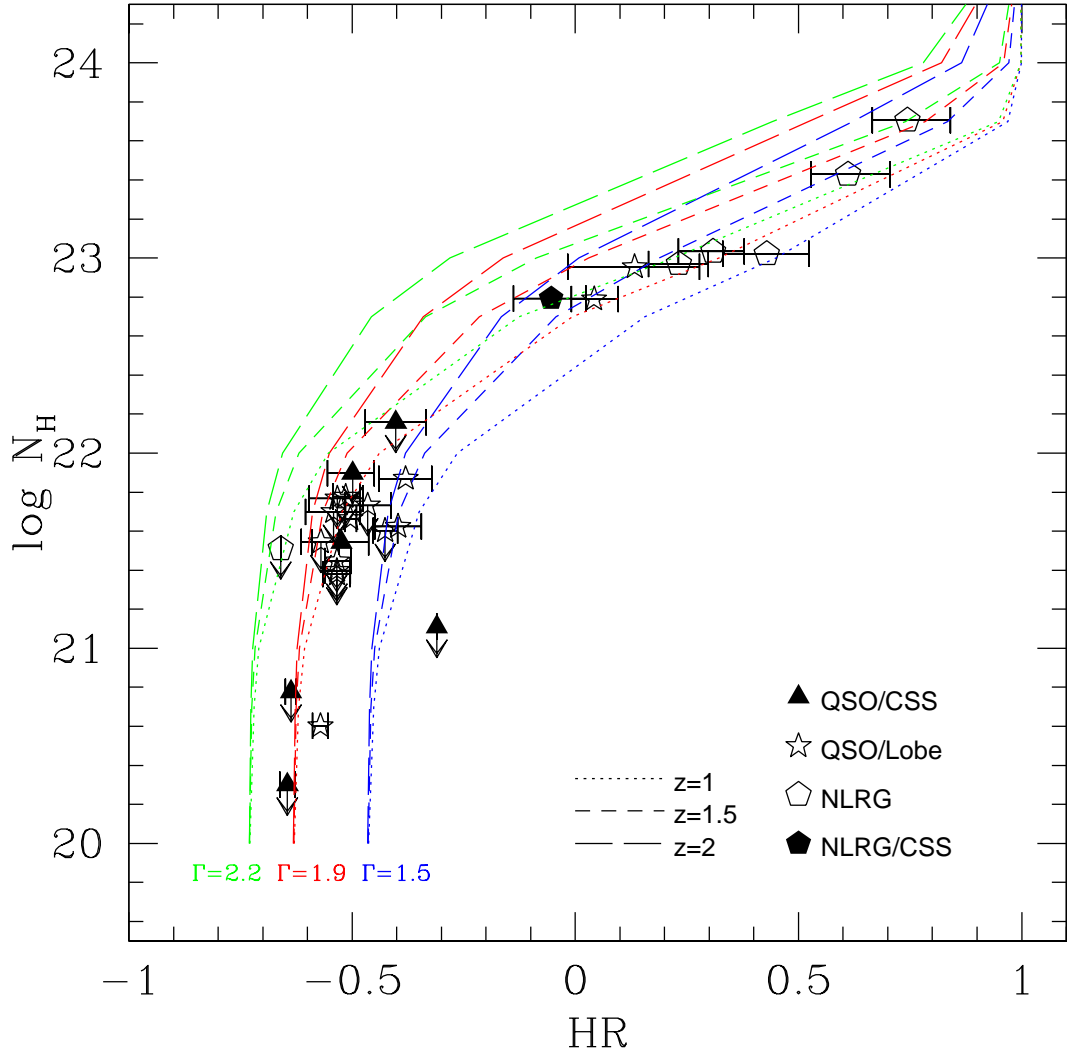}
\vspace{-0.5in}
\caption{The fitted \nh(int) (when available) as a function of 
the observed hardness ratio. For comparison, the relationship between
\nh(int) and HR for an absorbed power law is shown, assuming 
$\Gamma$ = 1.5(blue), 1.9(red), 2.2(green), at redshifts 1, 1.5, 2, 
ranges which cover the present sample.
The different symbols indicate the class of source as shown in the legend.
}
\label{fg:NHvsHR}
\end{figure}

The \nh(int) obtained from the spectral fits
is shown in Figure~\ref{fg:NHvsHR} as a function of 
the observed hardness ratio with lines showing the relationship between
\nh(int) and HR for a single, absorbed power law with several slopes and 
redshifts superposed for comparison.
The NLRGs trend similarly to models with the canonical 
quasar X-ray spectrum $\Gamma = 1.9$ \citep{2007ApJ...665.1004J}
with HRs indicating a maximum detected 
column density $\sim 7 \times 10^{23}$ cm$^{-2}$.
The fitted spectral slopes for the 7 quasars (non CSS) with $> 700$ net 
counts (Table~\ref{tb:flux}) show a mean of 1.69$\pm$0.05
and no evidence for a soft excess. This is harder
than the standard slope of 1.9,  consistent with a contribution
from beamed, jet-related emission which generally 
has a harder ($\Gamma \sim$1.5) 
slope (\citet{1987ApJ...323..243W}, see Section~\ref{sec:RLQ_X}).

\begin{figure}
\epsscale{0.8}
\plotone{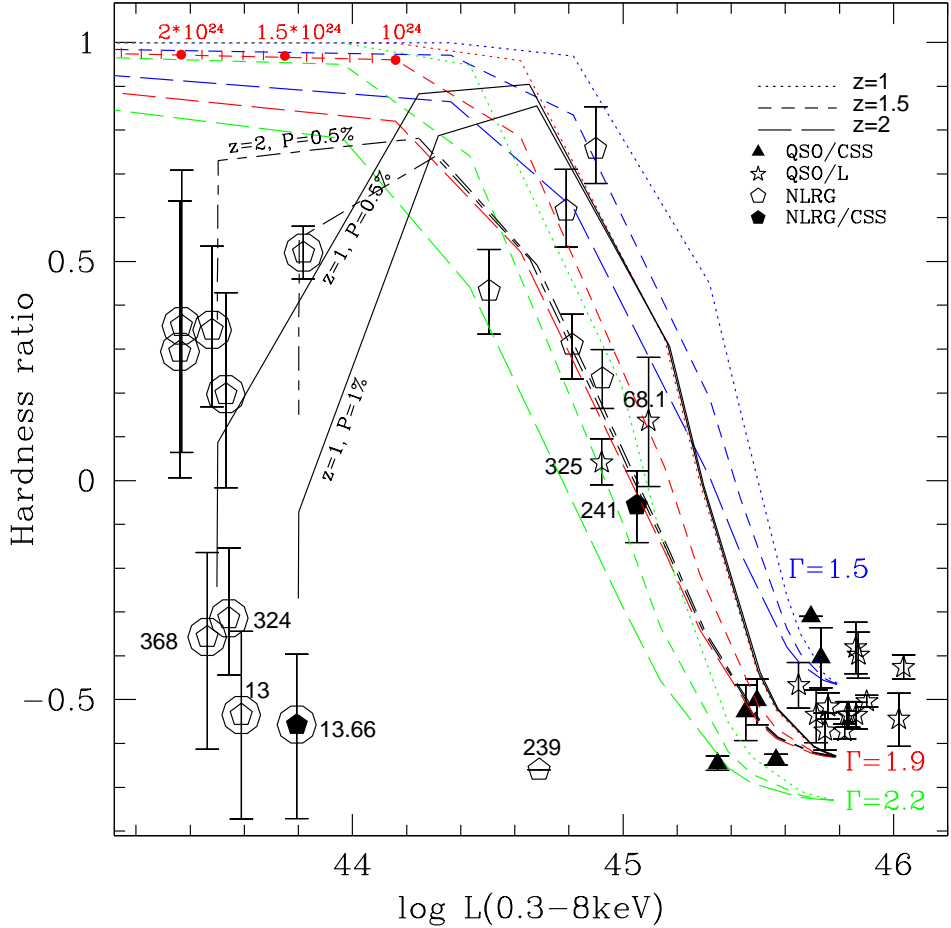}
\vspace{-0.5in}
\caption{X-ray hardness ratio as a function of broad band (0.3$-$8 keV)
X-ray luminosity in comparison with absorbed power law models 
assuming $\Gamma$ = 1.5(blue), 1.9(red), 2.2(green), at redshifts 1, 1.5, 2
(dotted, short-dashed, and long-dashed lines respectively), 
and covering \nh(int) = $1 \times 10^{20} - 1 \times 10^{25}$ cm$^{-2}$. 
\lx~is determined
without correcting for \nh(int) so as to demonstrate the effect of
absorption on the deduced \lx . Large circles indicate 
sources with $<50$ counts, for which \nh(int) could not be constrained
by the spectral fits. 
Red dots on the $\Gamma=1.9$, z=1.5 model curve (red short-dashed line)
indicate \nh(int) 
of (1,1.5,2)$\times 10^{24}$ cm$^{-2}$. 
The black lines show the addition of an unabsorbed 
power law ($\Gamma=1.9$) scattered at the 0.5 and 1\% levels 
to the z=1,2 models (solid and dashed black lines respectively) illustrating
one possible explanation for the softer spectra in the X-ray weakest sources.
}
\label{fg:LxvsHR}
\end{figure}

Figure~\ref{fg:LxvsHR} shows the hardness ratio as a function of the
broad band X-ray luminosity
determined from the spectral fit in our
standard analysis but uncorrected for any deduced \nh(int). 
Models for a power law
X-ray spectrum for a luminosity typical of the quasars in the sample
with a range of slope ($\Gamma$), intrinsic absorption (\nh(int)),
and redshift are shown for comparison.
The observed quantities are consistent with the models for X-ray
luminosities above $\sim 4 \times 10^{44}$ erg cm$^{-2}$ s$^{-1}$, where 
sources with lower luminosities than the quasars 
have harder HR as expected for mild absorption. However, as the 
luminosity decreases further 
the observed hardness ratios remain constant or soften.
This trend can be explained in terms of 
spectral complexity, where contributions from
weaker components (e.g. soft excess, reflection) become significant
as the dominant power law emission is absorbed away.
To illustrate the effect of a weaker component, the black lines in 
Figure~\ref{fg:LxvsHR} show the addition of nuclear power law emission
($\Gamma = 1.9$)
scattered from extended material at levels of 0.5\% and 1\% for 
z=1,2. This example demonstrates that an additional, soft component
can explain the HRs of the lowest \lx~NLRGs in this figure.

A more general measure of the relative X-ray luminosity is given by 
normalizing to the total radio luminosity.
Figure~\ref{fg:HRvsLxLr} shows the hardness ratio as a function of
X-ray to total radio luminosity ratio, \lxlr\ 
(=\lx (0.3-8 keV)/\lr ).
Since the range of \lx~for the quasars is small, the trends are very 
similar to those in Figure~\ref{fg:LxvsHR}. 
This figure clearly shows the 3 intermediate sources discussed earlier 
(quasars 3C~68.1/325 and NLRG 3C~241) which lie in
between the rest of the 
quasars and NLRGs suggesting intermediate obscuration levels.
The upwards arrows indicate the change in HR for soft NLRGs 3C~324/368
when using a smaller (1$''$ arcsec) circle to extract the counts,
excluding some of the extended emission clearly present in these two sources
(Table~\ref{tb:flux}, footnote 12).

For sources with $< 50$ counts, the spectral fits do not provide 
useful constraints on \nh(int). 
However, based on the comparison between models and data in 
Figure~\ref{fg:LxvsHR}, we conclude that the lower values of \lx~and 
\lxlr~in NLRGs are due to obscuration. 
Thus \lxlr~is a more reliable obscuration indicator than the 
X-ray HR which includes additional, soft X-ray emission components. 
The factor of $\sim 3 - 200 \times$  lower X-ray luminosities 
for the NLRGs indicate intrinsic column 
densities \nh(int) $\sim 5 \times 10^{22} - 2 \times 10^{24}$ cm$^{-2}$
in the current sample. 
This corresponds to \av $\sim 30 - 1000$, \citep{2000asqu.book..183S}.
Values of \nh(int) determined for the 9 low-count sources
(encircled in Figure~\ref{fg:LxvsHR}) using the models in that
figure are listed in Table~\ref{tb:f4nh}.

\subsection{X-ray and mid-IR Properties}
\label{sec:MIR}

\spitz~observations of the high-redshift 3CRR sample
\citep{2008ApJ...688..122H} 
demonstrate uniform power law plus silicate emission
SEDs for the quasars
while the NLRGs show a variety of SED shapes. The latter are
interpreted in terms of a range of quasar to host galaxy ratios and
absorption properties. Quasars and NLRGs separate well
in the optical depth of 9.7$\mu$m silicate absorption 
(\tauSi) with the NLRGs having higher values \citep{2010ApJ...717..766L}.
The level of \tauSi\ also tracks the source orientation 
(\rcd). The one exception in this sample is 3C~190, a CSS
quasar with significant silicate absorption 
(\tauSi = 0.60, see Section~\ref{sec:css}).

X-ray absorption column densities are often significantly
higher than those in the visible or IR 
(factors of 3-100, \citet{2001A&A...365...28M}), implying differing 
lines of sight, low gas-to-dust ratios perhaps due to high temperatures
in the material close to the nucleus, or a lack of small grains in the 
nuclear dust \citep{2004ApJ...616..147G}.
A reported correlation 
between the strength of the silicate absorption feature 
and estimated X-ray absorption \citep{2006ApJ...653..127S}
indicates $\sim 100\times$ higher X-ray (gas) column density. 
Figure~\ref{fg:IRvsXNH} shows the X-ray hardness (left) and
the estimated  \nh(int) (right) as a function of \tauSi~from
\citet{2010ApJ...717..766L}. 
Relative to an average quasar spectrum, which includes silicate 9.7$\mu$m 
emission, the NLRGs, including those with
soft hardness ratios, show significant \tauSi.
This implies that gas (X-ray absorption) and dust (IR absorption) are related.
However the intermediate sources, 3C~68.1/325,\footnote{The third, 
3C~241 was 
not observed with the \spitz~IRS} look like quasars
in the IR, with no significant \tauSi~and inconsistent
with a detailed gas/dust spatial correlation.  
Tests show a significant correlation of \tauSi\ with X-ray \nh(int)
in the current sample (P$<$0.0001, Kendall's $\tau$ test).

The three sources with the highest \tauSi\ (\gax 1) are well separated
from the remainder of the galaxies in Figure~\ref{fg:IRvsXNH}.
Of these, two X-ray soft NLRGs 3C~324/368 
are known to be located in edge-on host galaxies
\citep{1998MNRAS.299..357B,2010MNRAS.401.1500L}, supporting earlier
suggestions that the host galaxy contributes significantly to the IR 
obscuration in active galaxies and quasars 
\citep{2012ApJ...755....5G,2009ApJ...705...14D}.
The optical data for the third, NLRG 3C~469.1, are of too low quality to
confirm a similar edge-on view, 
but the extended, aligned radio and X-ray emission 
\citep{2010MNRAS.401.1500L} are suggestive.

Figure~\ref{fg:5to8vsHR} shows X-ray HR (left) and \nh(int) (right)
as a function of rest-frame \5to8 \footnote{k-corrected based on the observed 
slope between 8$\mu$m (IRAC) and 24$\mu$m (MIPS).} 
which is also
demonstrated to be an 
absorption indicator \citep{2008ApJ...688..122H}.
Quasars are relatively unabsorbed with high \5to8  
~while NLRGs
are absorbed with lower values. 
While the X-ray and IR absorption 
appear to be related, there is no significant correlation
with either X-ray HR or \nh(int).
The intermediate sources have quasar-like \5to8 . 
NLRG 3C~469.1 is once again unusual. It has very 
strong silicate absorption so that the derived rest-frame
\5to8  
~is unusually blue (Figure~\ref{fg:5to8vsHR}). 
Apart from this anomally, the IR and visible SED of 3C~469.1 
(Figure~\ref{fg:sed})
is red and consistent with that of a NLRG \citep{2010ApJ...717..766L}.

The near-IR obscuration of the NLRGs
deduced from the \5to8 ~ratio indicates an average obscuration 
\av $\sim50$ \citep{2008ApJ...688..122H},
while that indicated
by \tauSi~is lower, \av$\sim 20$ \citep{2010ApJ...717..766L}.
The X-ray data imply \av $\sim 30 - 1000$, again 
showing the tendency for the X-ray absorption to be higher.
The difference in the two IR absorption indicators suggests that the 
hot dust component ($\sim 4 \mu$m)
is closer to the AGN than the MIR emission/absorption 
region \citep{2011ApJ...729..108D}.

\subsection{Radio Core Fraction}

As noted earlier, at low frequencies all the 3CRR sources
are lobe-dominated, but at higher frequencies emission from the core
becomes significant. The relative strength of the core
emission (core fraction, \rcd) can be used as an orientation indicator.
Unification models predict a correlation between obscuration and orientation.
Figure~\ref{fg:LxLr_HRvsR}(left) shows \lxlr\ 
as a function of \rcd~with X-ray hardness ratio indicated by color.
A strong correlation is present (P$_{\rm null}$=0.0001 Generalized 
Kendall's $\tau$ test including upper limits on \rcd), consistent 
with Unification models (see also \citet{2005ApJ...634..169D} 
for lower-redshift, radio-loud AGN). 
The orientation dependence of the beamed, core emission 
is likely to also contribute to this relation. 
The significant relation between \nh(int) and 
\rcd~(P$_{\rm null}$=0.0001, Kendalls' 
$\tau$ test) shown in Figure~\ref{fg:LxLr_HRvsR}(right)) also strongly 
supports Unification models.


\section{Discussion}

\subsection{Compton Thick (CT) Candidates}
\label{sec:CT}

Four of the \chandra~observed
NLRGs, 3C~13/324/368, and 4C~13.66, have soft
X-ray hardness ratios consistent with a clear line-of-sight to a face-on
quasar while  their
X-ray luminosities are low, comparable to the other NLRGs. 
The count rates are sufficiently low that spectral fits provide no useful
constraint on intrinsic
column density for these sources (Table~\ref{tb:flux}). 
There are two possible interpretations. First, they could be 
LERGs which have little/no X-ray absorption 
and lack an actively accreting AGN (Section~\ref{sec:unif}), but since
the current sample is high luminosity this is unlikely.
Second, the direct AGN light in the soft NLRGs 
may be sufficiently absorbed (Compton thick, CT)
that the observed X-rays are dominated by other, relatively weak 
components, as observed in red quasars \citep{2009ApJ...692.1143K}
and consistent with the models in Figure~\ref{fg:LxvsHR}.

\begin{figure}
\epsscale{0.8}
\plotone{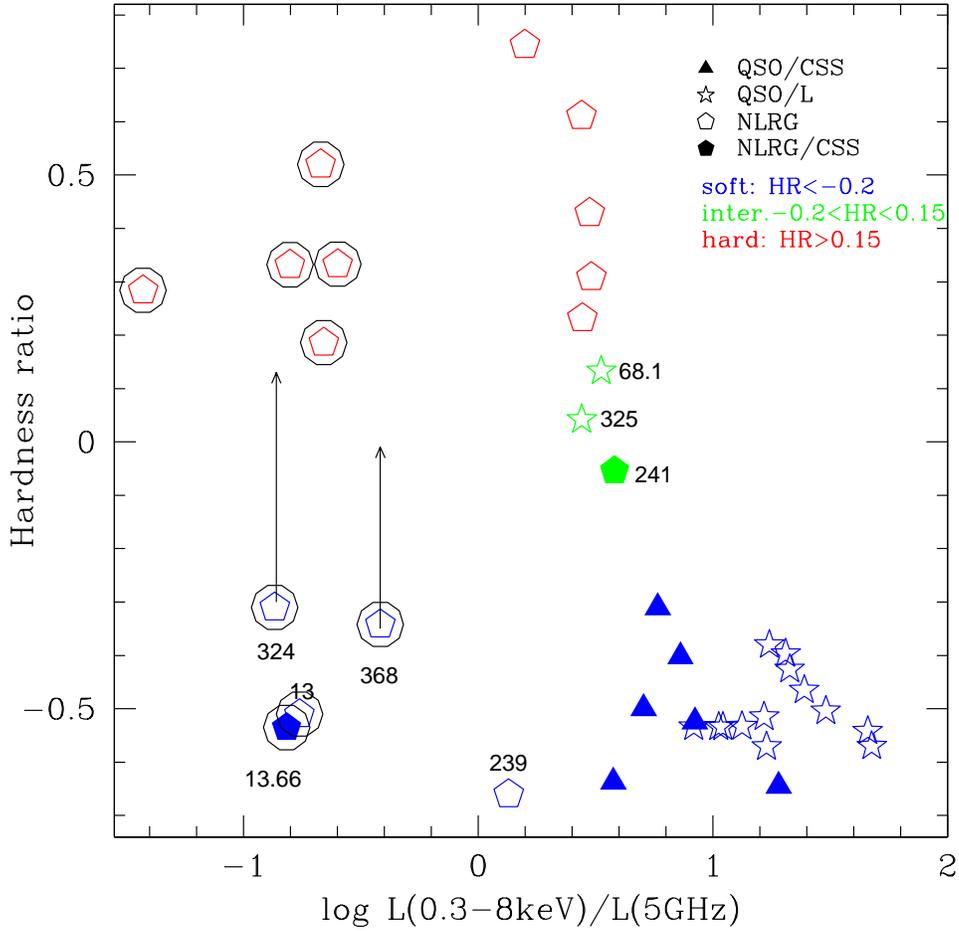}
\caption{X-ray hardness ratio (HR) as a function of 
X-ray to total 5 GHz radio luminosity ratio, \lxlr .
Symbol shapes and colors are indicated in the legend.
The upwards arrows for 2 soft NLRGs indicate the HR using a smaller (1$''$) 
extraction circle to exclude visible extended X-ray emission 
(Table~\ref{tb:flux} footnote 12).
}
\label{fg:HRvsLxLr}
\end{figure}

\begin{figure}
\epsscale{1.0}
\plottwo{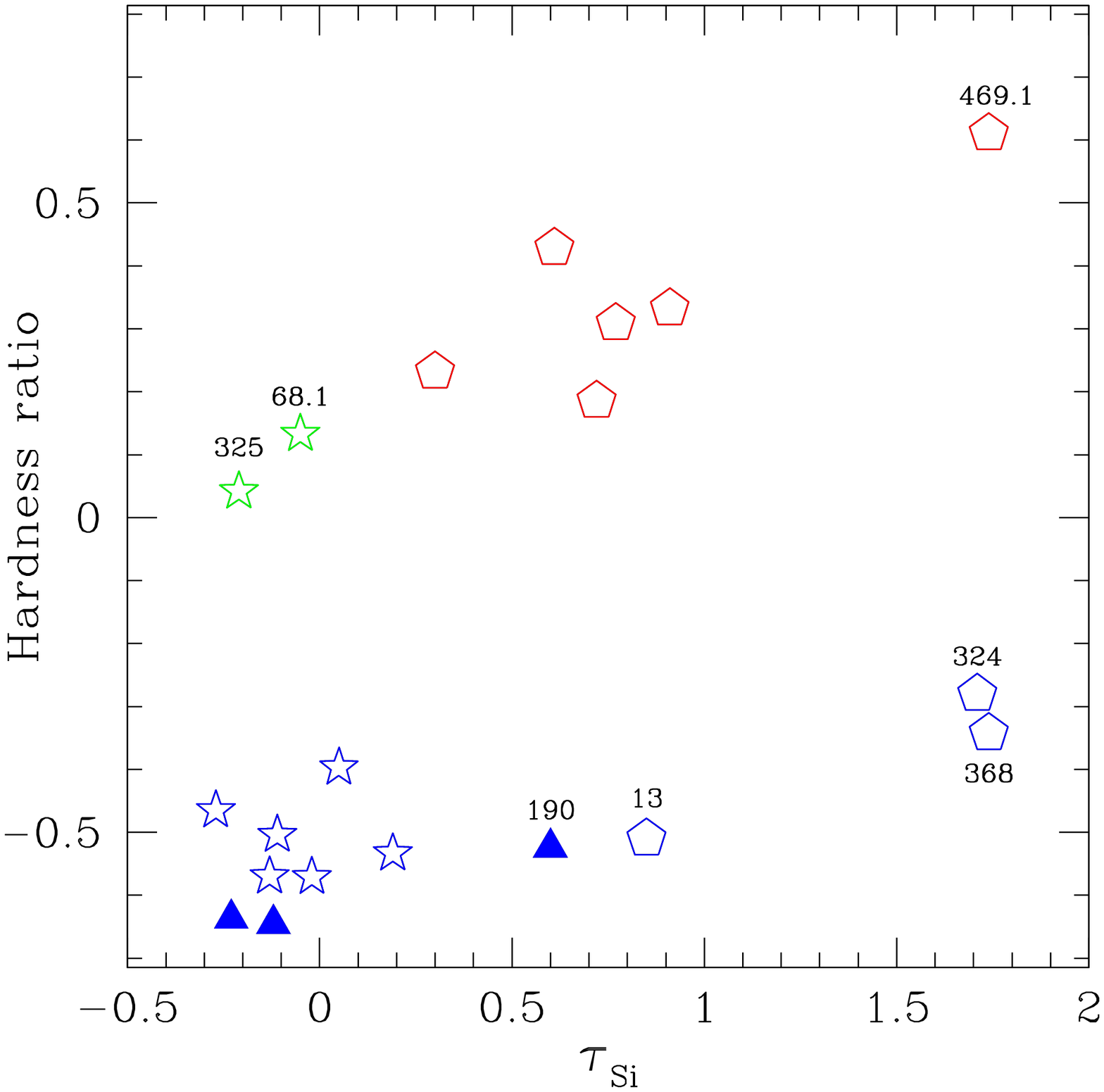}
{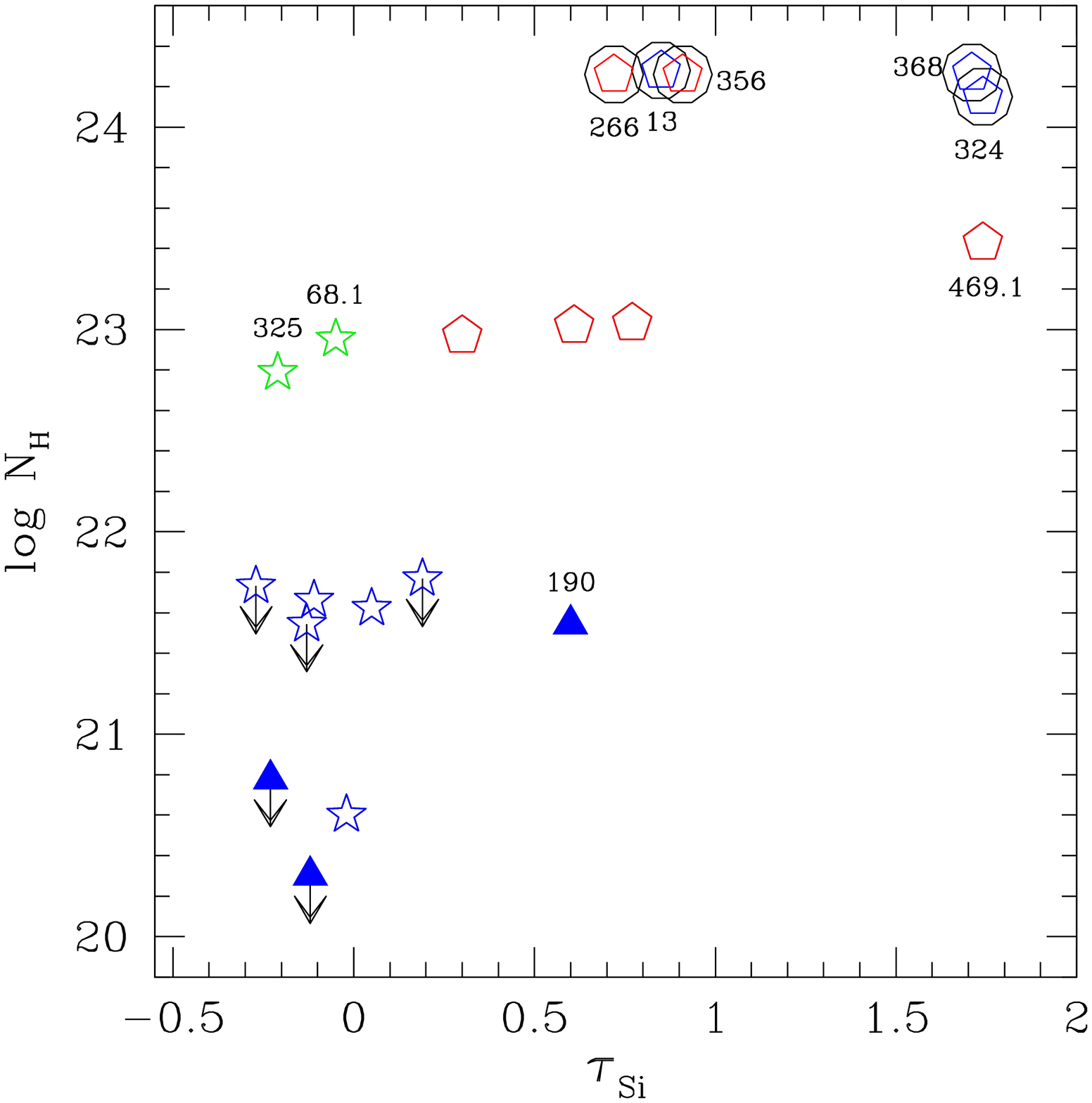}
\caption{The X-ray hardness ratio (HR, left) and the 
equivalent intrinsic hydrogren column density (\nh(int), right)
estimated from spectral fits or \lx~(circled data points)
as a function of the \spitz-measured optical depth of the silicate
$\lambda 9.7 \mu$m absorption \tauSi~\citep{2010ApJ...717..766L}. 
}
\label{fg:IRvsXNH}
\end{figure}

\begin{figure}
\epsscale{1.0}
\plottwo{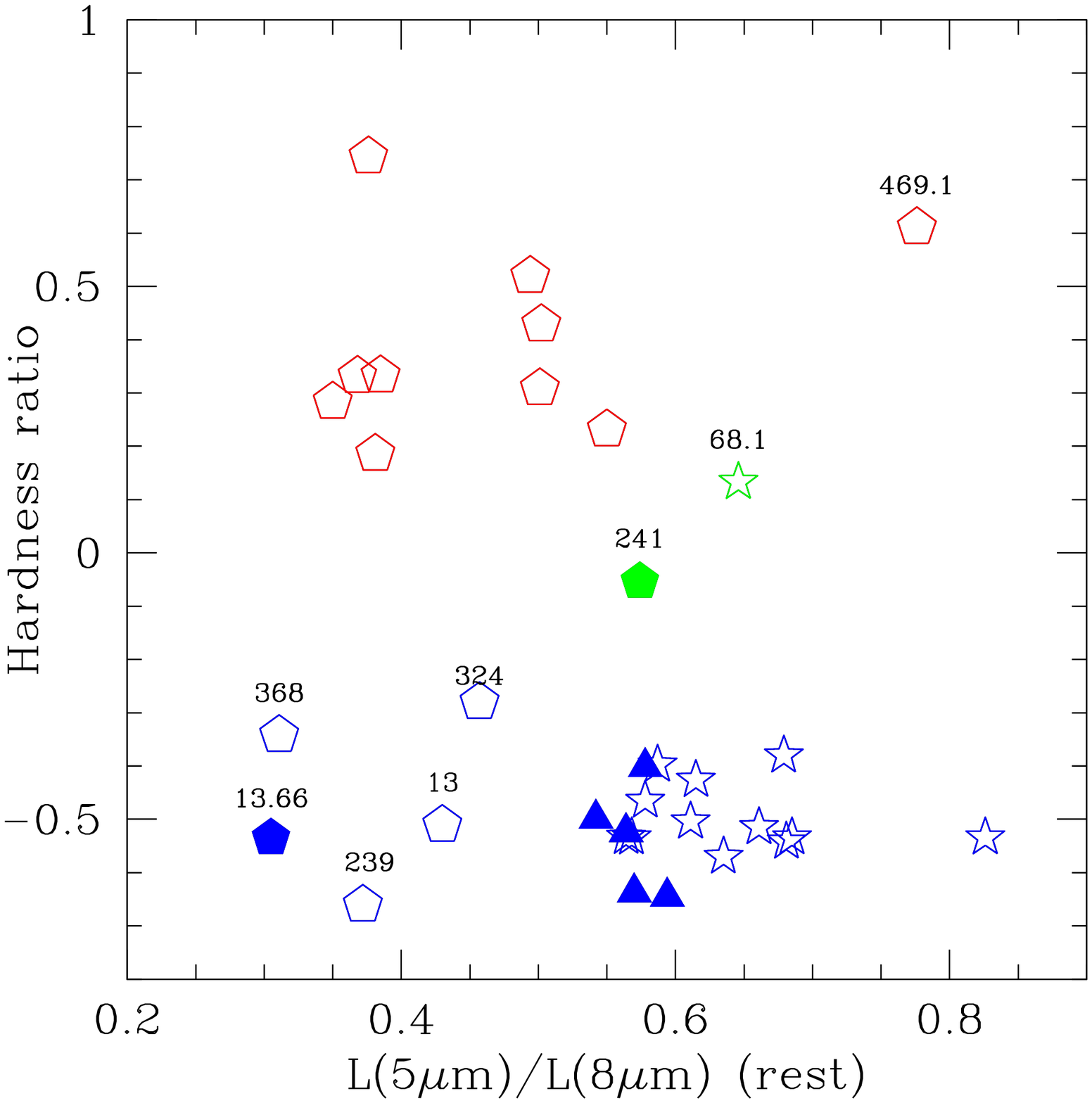}
{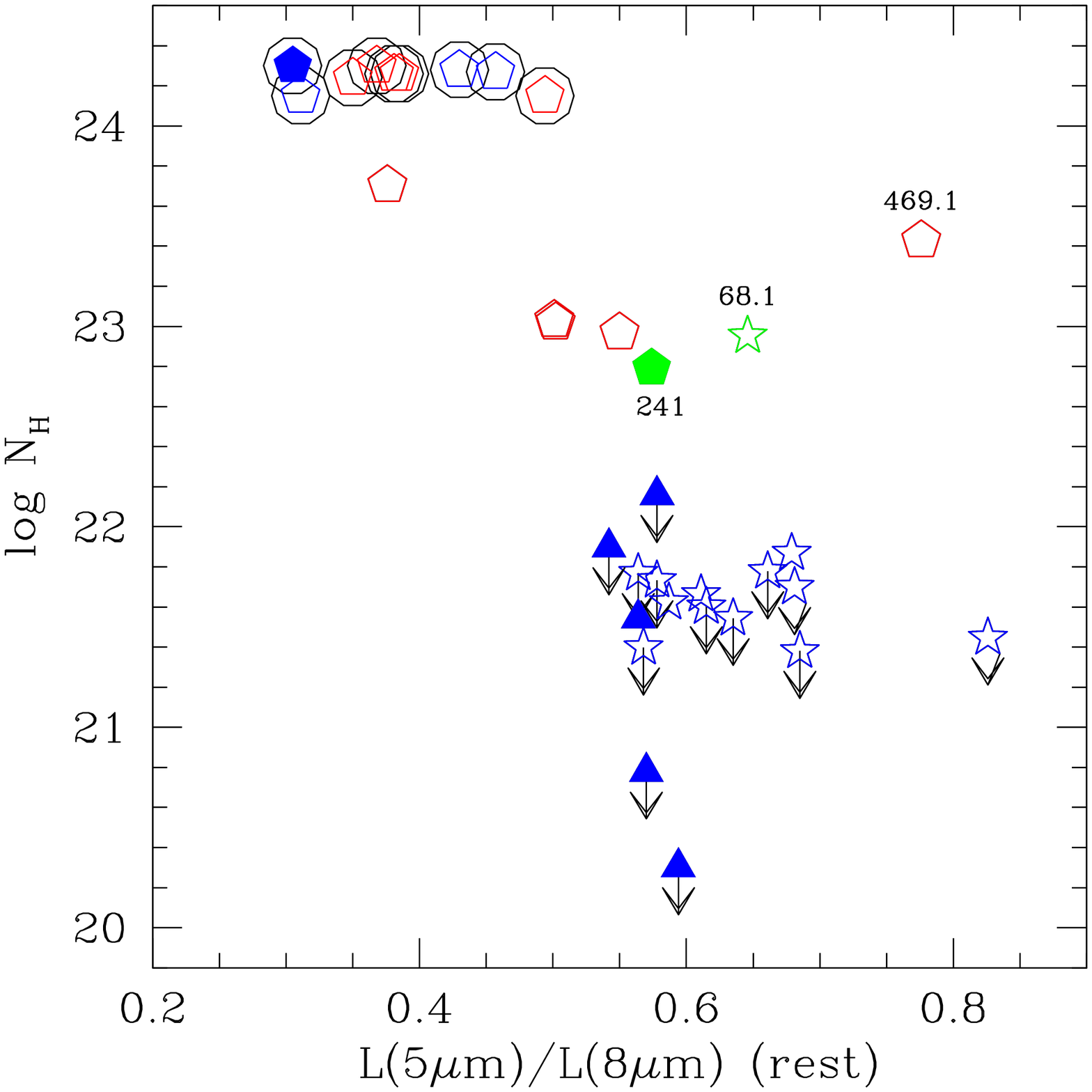}
\caption{The X-ray hardness ratio (HR, left) and the 
equivalent intrinsic hydrogren column density (\nh(int), right)
estimated from spectral fits or \lx~(circled data points) as a
function of the ratio of intrinsic 5$\mu$m to 8$\mu$m luminosities
(\5to8 ).
NLRG 3C~469.1 looks blue in \5to8 ~ because the deep \tauSi\ absorption affects
the observed 24$\mu$m band at redshift, z=1.336.
}
\label{fg:5to8vsHR}
\end{figure}

The circumnuclear regions of AGN are rarely resolved, particularly 
at high redshift, and so the extracted X-ray counts 
will include contributions such as
nuclear light scattered into our line-of-sight via dust/electrons above/below
the AD/torus, extended X-ray emission due to photo- or
collisional-ionization in material surrounding the nucleus, and/or
non-thermal emission from extended radio structure.
NGC 1068 is the archetypal CT source 
for which the X-ray emission was measured to be weak and soft in low 
spatial resolution data \citep{1987ApJ...315L..17M}. Detailed
\chandra\ and \xmm\ data have since revealed extended soft X-ray 
emission and a complex X-ray spectrum including multiple reflected components
\citep{2000MNRAS.318..173M,2006MNRAS.368..707P,2003A&A...402..849O}.
Although there are too few source counts in the soft NLRGs
to study source extent, counts were extracted from a smaller (1$''$ radius) 
circle to test this possibility. 
For 3C~324/368, the \chandra\ data show extended X-ray emission 
over a larger region and the resulting nuclear spectrum 
is harder (Table~\ref{tb:flux} footnote 12, Figure~\ref{fg:HRvsLxLr}), 
consistent with 
contamination by softer extended emission. The lack of similar 
spectral hardening for 3C~13 and 4C~13.66 does not rule out
the presence of unresolved extended soft emission.

The luminosity of the $[$OIII]$\lambda 5007$ emission line (hereafter 
L$[$OIII]) tracks the radio and X-ray luminosities for broad and 
narrow-lined AGN
\citep{1997MNRAS.286..241J,1994ApJ...436..586M}, and at high luminosities
there is little/no inclination dependence (see discussion in 
Section~\ref{sec:unif}).
Given the strong dependence of the observed X-ray flux on 
obscuration, L$[$OIII] is often used as an
indicator of the intrinsic X-ray luminosity, 
and the ratio of the two quantities indicates whether/not a source is CT
\citep{1999ApJ...522..157R,2006A&A...455..173P}.  
Figure~\ref{fg:O3LxvsR} shows the ratio L$[$OIII]/\lx \footnote{L$[$OIII]
measurements are from \citet{2004MNRAS.349..503G}. 
For 3C~43/204/325/437/469.1 
L$[$OIII] was determined from L$[$OII] using the relation reported
in that paper.}
as a function of \rcd\ for the 3CRR sample
in comparison with the CT criterion of \citet{2011ApJ...736..104J}.
Any weakening of the observed L$[$OIII] due to obscuration of the inner NLR
would imply a larger intrinsic L$[$OIII]/\lx .
All four of the \chandra-observed
X-ray soft NLRGs have high 
L$[$OIII]/\lx~ratios, consistent with CT X-ray emission. 
Five additional NLRGs 3C~68.2/266/294/356/437 
are also CT by this criterion. 
They have HRs $\sim 0.2-0.4$ (Table~\ref{tb:flux}) but their \lx~values are
comparable with the soft NLRGs and, in
comparison with the models (Figure~\ref{fg:LxvsHR}), indicate
\nh(int)\gax$1.5 \times 10^{24}$ cm$^{-2}$ (Table~\ref{tb:f4nh}).

Another parameter suggested to be a CT indicator is \tauSi\ 
\citep{2011A&A...531A.116G}. There are 3 sources with  
\tauSi $>$1: 3C~324/368/469.1 
(Figure~\ref{fg:IRvsXNH}, Section~\ref{sec:MIR}),
adding 3C~469.1 as a CT candidate. However
not all properties of this source align with a CT interpretation.
3C~469.1 has a relatively high \lx , a hard X-ray 
spectrum (Figure~\ref{fg:HRvsLxLr}),  
\nh(int)$\sim 3 \times 10^{23}$ cm$^{-23}$ (Table~\ref{tb:flux}),
and low L$[$OIII]/\lx~(Figure~\ref{fg:O3LxvsR}).
Although we cannot rule out that this source is intrinsically different, 
it seems most likely that
the nuclear absorption is not CT and that the unusually 
high \tauSi\ is dominated by the host galaxy.
Three other CT candidates 3C~13/256/266 have significant \tauSi $> 0.5$ 
while the remainder do not have the IRS data required to measure this feature.

The NLRG 3C~239, observed by \xmm, has a soft X-ray spectrum 
and possible Fe K$\alpha$ emission
leading \citet{2008A&A...478..121S} to suggest it is a CT source.
The \spitz~ \5to8  ~flux ratio aligns with the other soft NLRGs 
(Figure~\ref{fg:5to8vsHR}). However the radio
core fraction indicates an intermediate
orientation (Figure~\ref{fg:LxLr_HRvsR}),
\lx~is significantly higher than the other soft NLRGs
(Figure~\ref{fg:HRvsLxLr}),
and the low L$[$OIII]/\lx~ratio places it well inside the 
Compton-thin region (Figure~\ref{fg:O3LxvsR}).
Optical imaging shows a complex structure suggestive
of a merger remnant  \citep{1997MNRAS.292..758B}.
Given the conflicting properties and the relatively
low S/N of the \xmm~data,\footnote{such that the Fe K$\alpha$ emission
line is within 2$\sigma$ of 
the normal strength for a type 1 AGN, (Risaliti private communication)}
this source is not included in the list of CT candidates.
The unusual combination of soft X-ray spectrum and intermediate
X-ray luminosity (Figure~\ref{fg:HRvsLxLr}) suggest 
an intermediate level of absorption combined with
significant soft excess emission. Deeper X-ray data are
required to confirm/refute this suggestion.

The \chandra\ spectrum of the NLRG 3C~294 was studied at higher S/N 
by \citet{2003MNRAS.341..729F}, who report that it is dominated by a 
disk reflection component and deduce \nh=$(8.4\pm1) \times 
10^{23}$ cm$^{-1}$, within 2$\sigma$ of being CT, 
and an intrinsic \lx$\sim 1.1 \times 10^{45}$ erg cm$^{-2}$ s$^{-1}$. 
We thus consider the evidence that this source is CT to be marginal, 
and it is not included in our final CT list.

We conclude that there are 8 CT candidates amongst the
16 NLRGs (not including the intermediate source 3C~241 as an NLRG, Section 
\ref{sec:hardqs}) in this sample. We used the CIAO/Sherpa extension package 
Datastack to perform simultaneous X-ray spectral fits to the 8 datasets 
to explore their spectral form since individual fits provided no
constraints on the spectral parameters (Section \ref{sec:xrayanalysis}). 
Datastack allows the source redshifts 
and calibration files of the individual observations to be used appropriately
during simultaneous
fitting of a group of sources. The results confirm that 
no significant \nh(int) is detected and that a power-law provides a 
good fit ($\Gamma \sim 1.6$). However, since the total
number of counts for all 8 sources is 150, the fit does not provide 
strong constraints on the presence of a reflection component.

The 8 CT sources represent $\sim 21 \pm 7$\% of the 
3CRR sources in this redshift range.
Of these sources, 5 are well-documented aligned radio galaxies with 
extended optical/IR emission distributed along the
radio axis: 3C~13 \citep{1997MNRAS.292..758B},
3C~266 \citep{2003ApJ...585...90Z}, 3C~324 and 
3C~368 \citep{1998MNRAS.299..357B},
and 4C~13.66 \citep{1996MNRAS.279L..13R}, although this last is very small 
(1.4$''$) on the sky and the optical extension is not well-defined. 
The predominantly edge-on nature of these candidates for CT nuclear absorption
suggests that the host galaxy may contribute significantly to the 
absorption  
\citep{2012ApJ...755....5G,2009ApJ...705...14D,2009ApJ...692.1180K}.

\subsection{Intermediate Quasars and NLRGs}
\label{sec:hardqs}

Two of the quasars (3C~68.1/325) are unusually hard in the X-ray 
compared with the
remainder of the 3CRR quasars (Figure~\ref{fg:LxvsHR}). 
Their X-ray properties, along with those 
of NLRG 3C~241, are intermediate between those of NLRGs and quasars
(Section~\ref{sec:xrayabs}).

\begin{figure}
\epsscale{1.0}
\plotone{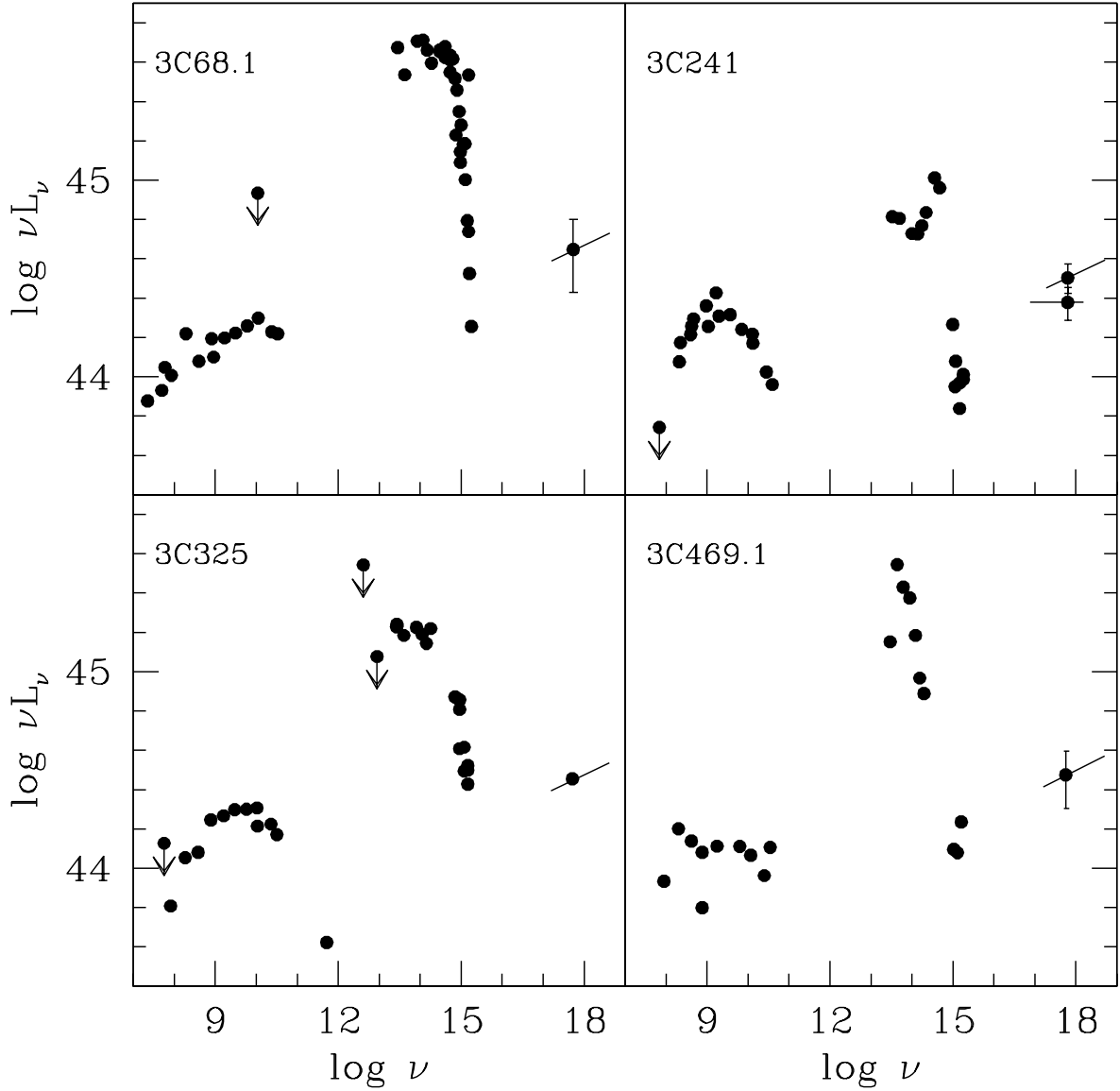}
\caption{Radio$-$X-ray spectral energy distributions (SEDs) of the intermediate
QSO/NLRG sources as labelled in each panel and the unusual source 3C~469.1. 
Upper limits are indicated at 3$\sigma$ and the X-ray points indicate
the estimated spectral slopes.
All four sources have a red optical-UV continuum with 
little evidence for a blue bump indicating strong UV absorption. The lack of 
visible galaxy emission is consistent with the observed 
moderate X-ray absorption, \nh$\sim 10^{22-23}$ cm$^{-2}$.
3C~469.1 also has strong \tauSi~absorption.}
\label{fg:sed}
\end{figure}

\begin{figure}
\epsscale{1.0}
\plottwo{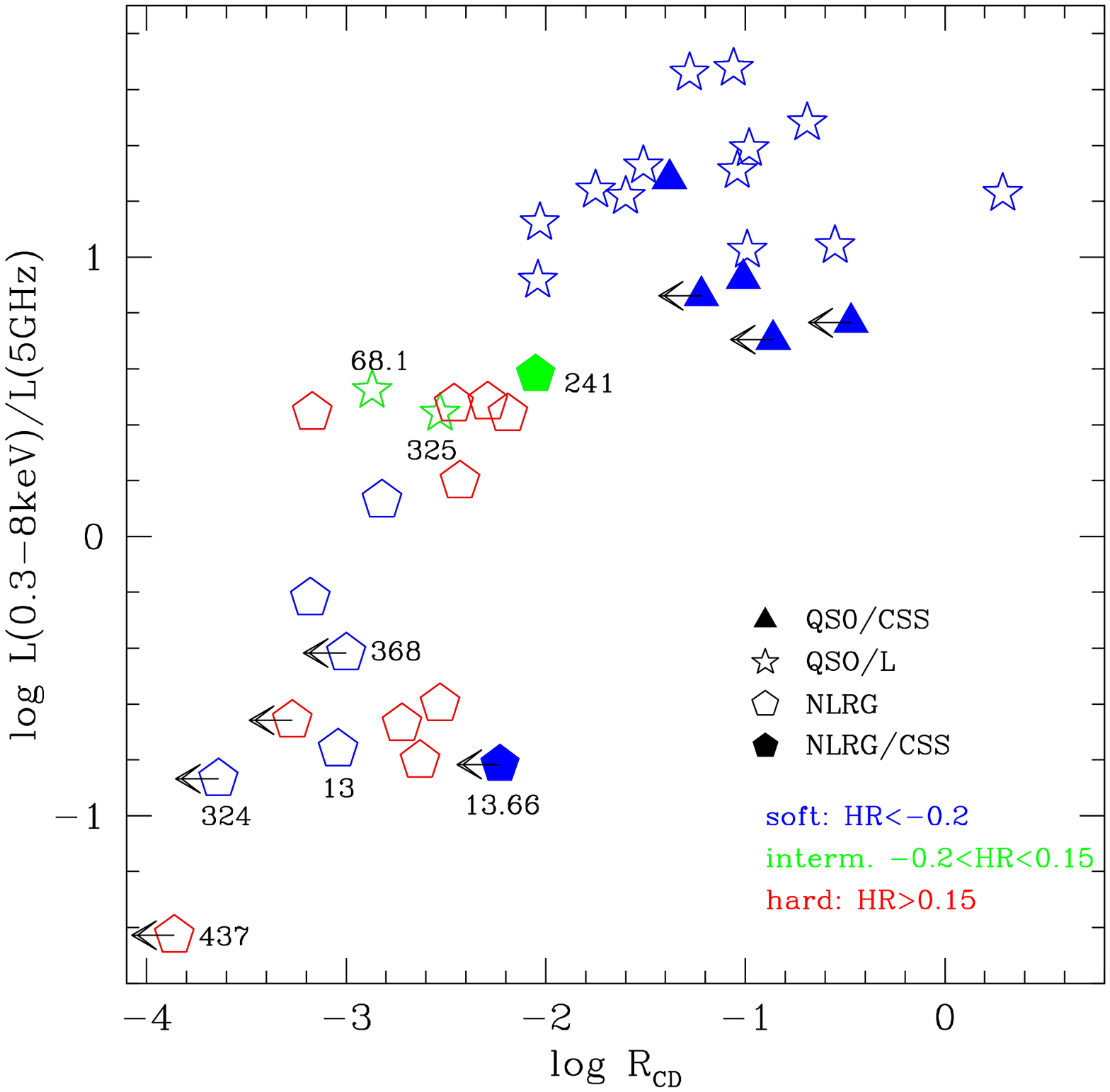}
{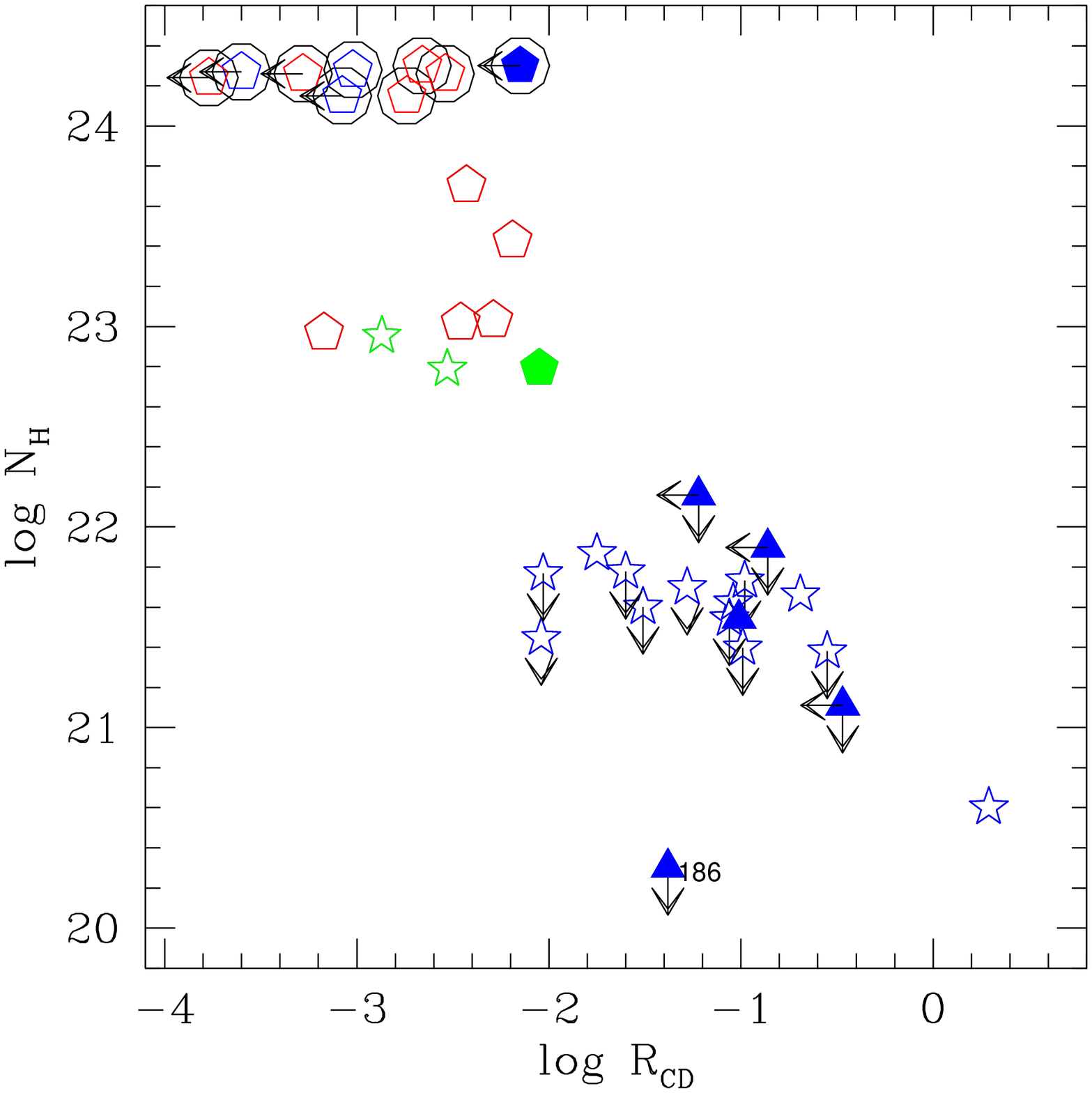}
\vspace*{-0.5in}
\caption{The X-ray to total radio luminosity ratio (\lxlr, left) and 
intrinsic equivalent hydrogen column density (\nh(int), right) estimated 
from spectral fits or \lx~ (circled data points) 
as a function of the radio core fraction \rcd. 
A strong relation with \rcd~is present for both parameters 
consistent with the orientation-dependent obscuration
of Unification models.
}
\label{fg:LxLr_HRvsR}
\end{figure}

\begin{figure}
\epsscale{0.8}
\plotone{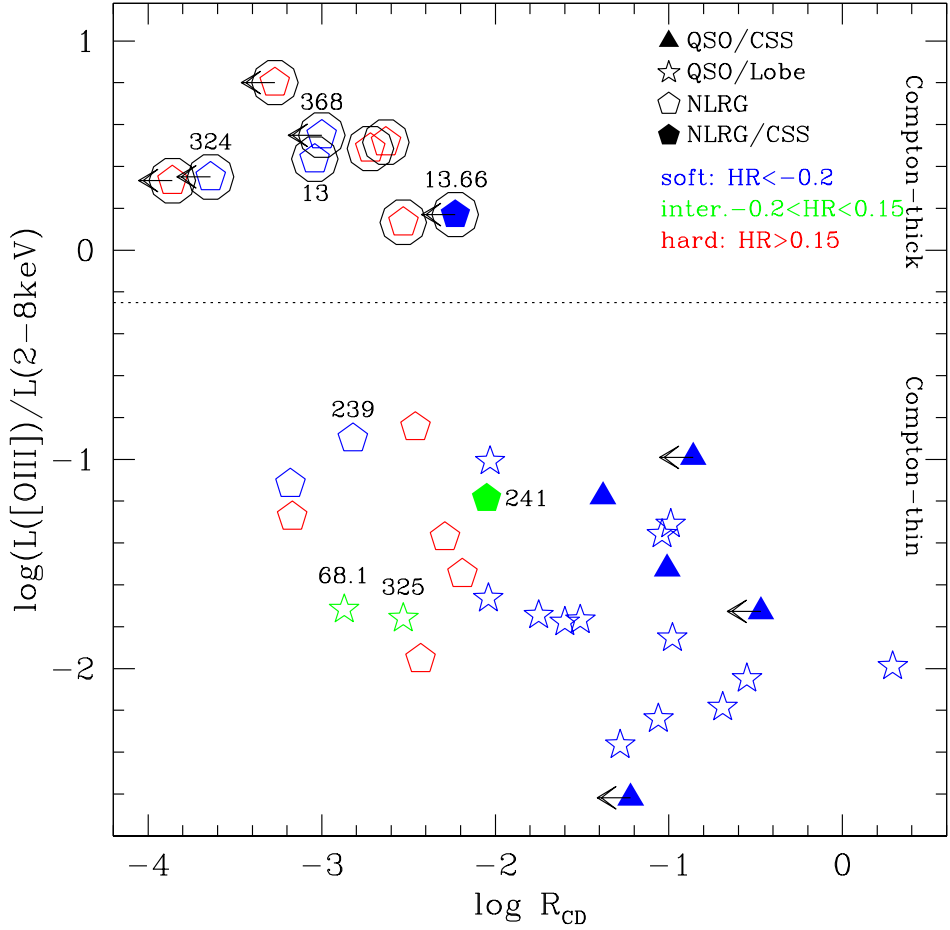}
\vskip -1.0in
\caption{The ratio of 
L$[$OIII] to hard (2$-$8keV) X-ray luminosity 
(not corrected for \nh(int)) as a function of radio core fraction \rcd . 
Symbols and colors are indicated in the legend.
For five sources with no measured L$[$OIII]
(3C~43/204/325/437/469.1) values
were estimated from measurements of L$[$OII] following 
\citet{2004MNRAS.349..503G}.
Nine NLRGs lie in the range of L$[$OIII]/\lx~expected 
for Compton thick (CT) AGN as
indicated by the dotted line \citep{2011ApJ...736..104J}.
}
\label{fg:O3LxvsR}
\end{figure}

3C~68.1 has hard X-ray emission, while the silicate absorption at 9.7$\mu$m 
is weak and in the quasar range \citep{2010ApJ...717..766L}.
The SED (Figure~\ref{fg:sed}) shows a red optical/near-IR continuum
and little/no blue bump.
The optical spectrum includes broad and narrow 
emission lines and strong, narrow
absorption close to the emission line redshift (v$\sim -70$ km s$^{-1}$, 
\citet{1998ApJ...501..110B}). The optical continuum and broad lines are
highly polarized, ranging from $\sim
5-10$\% progressing from red to blue along the spectrum.
The combination of high polarization, no strong
variability, broad lines, weak (undetected in the ROSAT All-Sky
Survey) X-rays 
and strong UV absorption was interpreted  by \citet{1998ApJ...501..110B}
as due to an inclined
system where the scattered, polarized emission is diluted by dust-reddened
direct light towards the red end of the optical spectrum. The
\chandra\ detection shows relatively weak
X-ray emission obscured by an intermediate absorbing column density 
(\nh(int) $\sim 9 \times 10^{22}$ cm$^{-22}$, Table~\ref{tb:flux}), 
confirming this picture.

3C~325, originally classified as a NLRG, was
re-classified as a quasar based on the presence of weak broad
components to the optical emission lines \citep{2005MNRAS.359.1345G}.
That paper also updated the redshift to 1.135 (earlier reported to be 0.86).
The \spitz~IR data confirm the re-classification, showing the strong, 
smooth, power-law-like continuum and silicate emission that are
characteristic of the 3CRR quasars \citep{2010ApJ...717..766L}.
The SED (Figure~\ref{fg:sed}) shows
a red optical/near-IR continuum and little/no blue bump.
The \chandra~ data show relatively weak X-ray emission
and intermediate absorption column density 
(\nh(int)$\sim 6 \times 10^{22}$ cm$^{-2}$, Table~\ref{tb:flux}).

3C~241 is classified as an NLRG/CSS, but a broad H$\alpha$ line has been 
observed in this source \citep{2003MNRAS.346.1009H}
implying that it is also intermediate between quasars and NLRGs. 
The SED (Figure~\ref{fg:sed}) shows a red optical/near-IR continuum and
$\sim$no blue bump. There are no \spitz~IRS data for this source.
The \chandra~data again show
relatively weak X-ray emission and intermediate 
absorption column density
(\nh(int)$\sim 6 \times 10^{22}$ cm$^{-2}$, Table~\ref{tb:flux}).

The multi-wavelength properties of all 3 sources
are very similar to those of red AGN and suggest type 1 quasars
with an intermediate level of obscuration 
so that the X-ray flux remains relatively strong 
\citep{2002ApJ...564L..65W,2005ApJ...634..183W}. The
AGN IR bump outshines the host galaxy emission but
the optical/UV emission is largely obscured \citep{2009ApJ...692.1143K}.

Numerous hybrid sources whose classification as type 1 or 2 AGN depends on the
observed waveband have been reported over the years. Galaxies bright
in the IR and dominated by a starburst may contain an AGN. AGN
with narrow lines in the optical reveal broad lines in IR observations
or in polarized optical light. Specific AGN classes include XBONGs
(X-ray bright, optically normal galaxies, 
\citet{2005MNRAS.358..131G}) 
or optically dull AGN \citep{1981ApJ...246...20E}, red quasars (e.g. 
2MASS sample, \citet{2002ASPC..284..127C}), type 2 quasars 
\citep{2008AJ....136.1607Z,2006ApJ...637..147P}.
Explanations include an evolutionary stage in which a quasar 
is emerging from an early, enshrouded state 
\citep{2006ApJS..163....1H,1988ApJ...325...74S},
an intermediate orientation of the AGN in which the
quasar is viewed through a lower column density of material and/or an
edge-on host galaxy\citep{2009ApJ...692.1143K,
2009ApJ...692.1180K,2002ApJ...564L..65W,2000ApJ...545...63E}, 
dilution by a bright and/or
edge-on host galaxy, or an intrinsically weak AGN 
\citep{2009ApJ...706..797T,2009MNRAS.398..333H}. Since this is a luminous, 
radio-selected sample, 
the last two possibilities, which imply  a weak AGN, seem  unlikely. 
While an evolutionary stage with a weak AGN cannot be ruled out,
we will see below (Figure~\ref{fg:LxLr_HRvsR}) that the core fraction of
these three sources supports an intermediate
viewing angle with a lower column density of obscuring material.
This would be identified with a corona/wind above or below the 
AD/torus
as posited in current models \citep{2000ApJ...545...63E,1994ApJ...434..446K}
or a relatively clear line-of-sight 
in a clumpy torus model \citep{2008ApJ...685..160N}.

\subsection{Compact Steep Spectrum (CSS) Sources}
\label{sec:css}

The 3CRR sample includes 8 CSS sources: 6 are classified as quasars,
1 as a NLRG (CT candidate 4C~13.66), and 1 as an intermediate source (3C~241).
The radio source size is smaller than a typical galaxy,
and models generally involve a young radio source in a later
stage than the GHz peaked sources (GPS), which tend to be smaller
with radio structure comparable in size to the NLR \citep{1998PASP..110..493O}.
The lack of X-ray absorption in \chandra\ observations 
of both GPS and CSS sources \citep{2008ApJ...684..811S}
ruled out earlier models in which the radio source was confined by
interaction with a surrounding medium.
Their X-ray properties are consistent with 
unification: the quasars are not heavily obscured while the NLRGs are.
The small radio size then indicates that they
are young rather than confined by an external medium 
whose presence would be clear from additional X-ray absorption
\citep{2008ApJ...684..811S}. However, comparison of the X-ray properties
of a well-defined sample of GPS sources with a heterogeneus sample
of radio galaxies and quasars suggests that GPS sources are X-ray weak
(by a factor $\sim 10$) and somewhat obscured \citep{2009A&A...501...89T}.

Our sample provides a well-matched set of CSS and other radio sources.
Their X-ray properties generally align with those of the rest of the sample. 
The quasars show no absorption, the NLRG 4C~13.66 is a CT candidate and 
the intermediate source 3C~241 
is similar to the other two in that category (Figure~\ref{fg:LxvsHR}).
Figure~\ref{fg:HRvsLxLr} shows that the CSS 
quasars are at the low end of the \lxlr\ range for quasars. The mean ratio 
for CSS quasars is a factor of 2.5 lower than that for 
the rest of the quasars due to a 
combination of higher \lr\ and lower \lx . 
The fitted X-ray slopes for the two sources with sufficient counts
(Table~\ref{tb:flux}) are relatively soft: $\Gamma \sim 1.8$ (3C~287), 
$\Gamma \sim 2.1$ (3C~186, see also 
\citet{2008ApJ...684..811S}) compared with the 
other quasars (mean $\Gamma \sim 1.7$).  Thus the X-ray properties of
this uniform set of CSS quasars support earlier conclusions that the
radio sources are young rather than confined. The CSS quasars
have \lxlr~values a factor of $~\sim 2.5$ lower than the well-matched
quasars in this sample,
a smaller shift than that reported by \citet{2009A&A...501...89T}.
The combination of lower \lxlr\ and softer X-ray slopes
could result from weaker jet-related X-ray emission, 
but a larger well-matched sample is needed to confirm a systematic 
discrepancy.

3C~190 is a CSS quasar with conflicting properties. The X-ray properties
align with the other CSS quasars, but the IR data indicate 
silicate absorption (\tauSi$\sim 0.6$, \citet{2010ApJ...717..766L})
rather than the typical emission. 
This measurement was made with respect to the silicate emission from 
an average quasar so that 
an alternative interpretation is of weak/absent silicate emission.
It has been suggested (\citet{2010ApJ...717..766L} and Ogle et al.,
in preparation) that CSS quasars may have
unusual silicate features, although the other CSS quasars in the current
sample have silicate emission similar to normal quasars, not supporting
this idea.
Other possibilities include significant host galaxy absorption 
or a complex X-ray spectrum so that the deduced low \nh(int) is
incorrect.

\subsection{The Distribution of Intrinsic X-ray Absorption Column Densities}

\begin{figure}
\epsscale{0.8}
\plotone{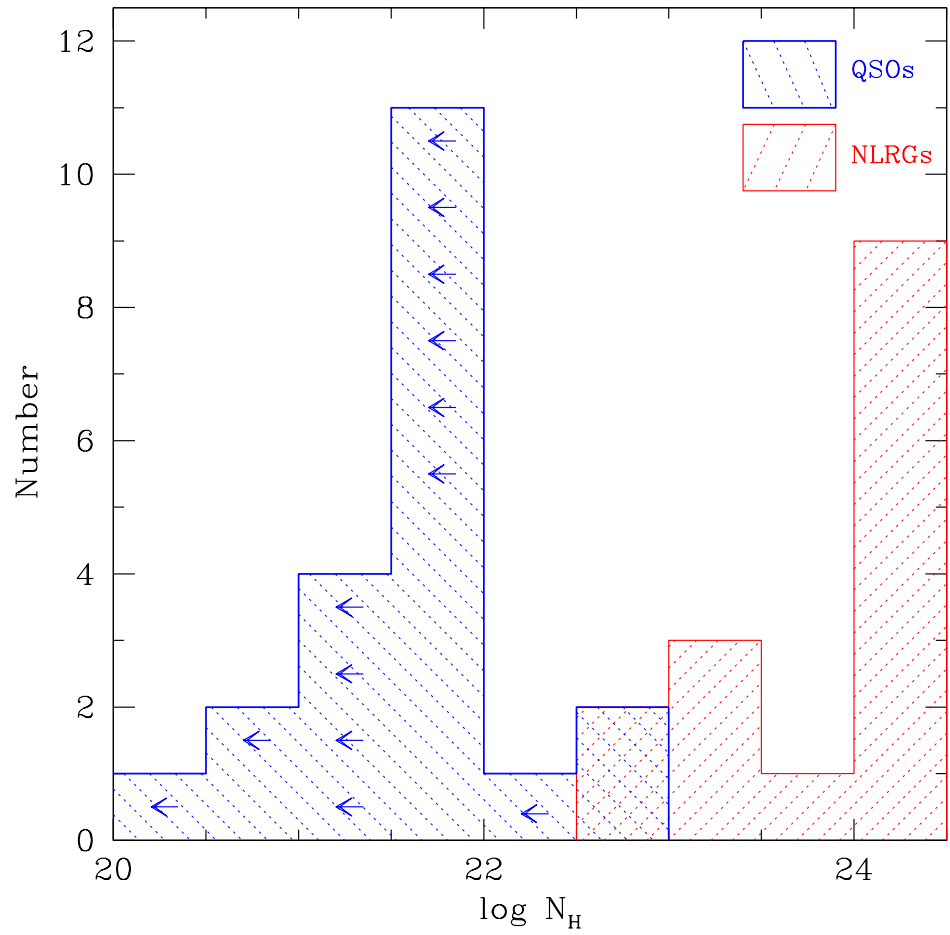}
\vskip -1.0in
\caption{The distribution of the best estimates of X-ray equivalent intrinsic 
hydrogen column density 
for the \chandra-observed sample. Quasars are shown in red and NLRGs in blue 
with upper limits, mostly for quasars with no evidence for intrinsic 
absorption, indicated by arrows.
}
\label{fg:NHdist}
\end{figure}

The best estimates of the intrinsic absorption column densities, \nh(int) 
for the 3CRR sources (Tables~\ref{tb:flux}, \ref{tb:f4nh}) 
were determined with reference to both X-ray and multi-wavelength properties
of each source (Section~\ref{sec:xrayabs}).
The resulting \nh(int) distribution is bi-modal (Figure~\ref{fg:NHdist}) 
with \nh(int) for the NLRGs peaking at $> 10^{24}$ cm$^{-2}$. 
A similar distribution, with obscured sources 
peaking $> 3 \times 10^{23}$ cm$^{-2}$,
was reported for lower-redshift 3CRR galaxies (z$<$1,
\citet{2006ApJ...642...96E,2009MNRAS.396.1929H}, their Figure 16).
In that case, the \nh(int) was based on spectral fits which include 
two power law components, one of which (accretion-related) is
absorbed while the other (radio-jet-related) is not. 
The current \chandra\ data are not of sufficiently high S/N to 
allow the multi-component fitting used in the low-redshift sample.
Despite the different analysis, the similarity of the two distributions 
reinforces the earlier conclusion
that the \nh(int) distribution for radio-selected AGN extends to higher values
than that of type 2 sources found in the SDSS. Distributions derived 
for lower-luminosity, local Seyfert 2 galaxies are more similar
\citep{1999ApJ...522..157R}. Once again this 
result emphasizes that low-frequency radio selection includes the 
full population,
including the highly-obscured (edge-on) sources that selection in other 
wavebands preferentially misses.

Low S/N observed X-ray spectra do not accurately reflect the true level of 
obscuration for highly-obscured sources. Even in the high-luminosity 
3CRR sample, some objects display weak and soft X-ray emission because
the primary X-ray power law is sufficiently obscured that weaker
X-ray components dominate. 
Using HRs or simple power law fits to estimate
the absorption column densities may
yield values $10-1000 \times$ too low and lead to
intrinsic X-ray luminosities $\sim 10-100\times$ below the true values. 
Underestimated absorption would result in an apparent lack of
heavily obscured sources in X-ray-selected AGN samples because such sources
either appear to be unabsorbed, adding to the unobscured rather than 
the obscured source counts, or they fall below the flux limit and are missing
from the sample altogether.

From the distribution of \nh(int) we conclude that the 
obscured fraction in the high-z 3CRR 
sample is 0.5$\pm$0.1, and the CT fraction is 0.21$\pm$0.07, both
consistent with CXRB model predictions in this range of 
\lx~\citep{2007A&A...463...79G}. This obscured 
fraction is higher than is typically
reported for sources in the same luminosity range ($\sim 0.1-0.3$, 
see Section~\ref{sec:unif}), a discrepancy which can be explained
if the $\sim 20$\% CT sources are generally either undetected
or accounted as unobscured.
Similar CT fractions ($\sim$0.18) have been reported for z$>3$, 
lower \lx~(log \lx~$\sim 43-44$) X-ray-selected sources in the CDFS 
\citep{2012A&A...537A..16F} and in low-redshift (z$<0.1$)
hard X-ray surveys \citep{2011ApJ...728...58B}. In summary, the ratio of
unobscured to Compton-thin
($10^{22}<$ \nh(int) $< 1.5 \times 10^{24}$ cm$^{-2}$)
to CT (\nh(int)$> 1.5 \times 10^{24}$ cm$^{-2}$) is 2.5:1.4:1
for the high-z 3CRRs.

\subsection{Geometry of the Nuclear Region}
\label{sec:Geom}

The strong relations between the X-ray-to-radio luminosity ratio (\lxlr)
and the intrinsic absorption column density (\nh(int))
and the AGN orientation as indicated by \rcd~(Figure~\ref{fg:LxLr_HRvsR})
confirm that orientation-dependent obscuration dominates the nuclear X-ray 
and core radio properties of high-redshift RLAGN, supporting the 
Unification model.
Anomalously high \tauSi~in a few sources indicates that the host galaxy 
makes significant contributions to the SEDs when the spatial 
resolution is insufficient to isolate the nuclear emission.

Estimates of the number of sources
as a function of the amount of obscuration provide constraints on the
covering factor of the material in the nuclear regions immediately 
surrounding the central SMBH. Half of the sample (19 of 38) 
are type 1 sources (obscured fraction of 0.5) 
with little/no X-ray absorption, strong, broad emission 
lines, and blue visible colors. 
This is consistent with previous estimates for samples of 
radio galaxies once the LERGs are removed
\citep{1989ApJ...336..606B,2006ApJ...647..161O}.
Assuming randomly-oriented 3CRR sources and a geometry in which 
the obscuring material 
lies preferentially in a plane perpendicular to the radio jet,
the probability of a source lying in a cone of angle $\phi$
is given by P($\theta < \phi$) = 1$-$cos$\phi$ \citep{1989ApJ...336..606B}
leading to an estimated half-opening angle for the obscuring material of 
60$\pm 8$\deg . 

The 3CRR sample includes 8 Compton-thick candidate NLRGs with one/more of the 
following properties:  low X-ray luminosity (\lxlr, sometimes accompanied by 
soft X-ray spectra), 
high $[$OIII]$\lambda$5007 to X-ray luminosity ratio, low \rcd, high \tauSi.
In a Unification scenario, these are the highest inclination sources,
viewed through the optically-thick material of the AD/torus
so that only weaker emission, reflected/scattered from cold and/or warm 
(ionized) material outside the nucleus, is visible in the X-rays
i.e. similar to the archetypal edge-on AGN NGC 1068.
The CT candidates represent $50 \pm 18$\% (8 of 16) of the NLRGs in 
this sample.
Six of these sources are also in edge-on or merging 
host galaxies suggesting that host galaxy obscuration also plays a role.
With $21 \pm 6$\% of the total sample in this category, we estimate 
that the CT
torus/AD covers $12 \pm 4$\deg\ above and below the equatorial plane of the
system.

The remaining 8 NLRGs and the 3 intermediate sources\footnote{The optical 
and IR spectral data for many of the NLRGs, which are very faint 
optical sources, are of low S/N making weak broad lines difficult
to detect so that 
the NLRG vs intermediate classification is non uniform.} 
(Section~\ref{sec:hardqs}), have Compton-thin \nh(int) 
and intermediate \lxlr~and \rcd. This group constitutes
$29 \pm 7$\% of the sample, indicating obscuration above 
and below the CT disk covering a further $18 \pm 3$\deg.
The presence of
3 intermediate class sources suggests that the material decreases in 
density away from the plane of the disk, perhaps including an
atmosphere/wind \citep{1994ApJ...434..446K,1995ApJ...454L.105M}.


\section{Conclusions}

\begin{enumerate}
\item \chandra\ X-ray observations of a complete
sample of 38 high-redshift (1\lax z\lax2)
3CRR radio sources (log \lr $\sim 44-45$, log \lx $\sim 43-46$)
combined with multi-wavelength data have demonstrated that 
source orientation can explain the full range of X-ray properties, consistent
with the orientation-dependent obscuration of the
Unification model for radio-loud quasars and radio galaxies (NLRGs).

\item The obscured fraction for this sample of high-redshift 3CRR sources 
is 0.5$\pm$0.1, consistent with that expected at log \lx $\sim 45-46$ 
in CXRB models \citep{2007A&A...463...79G}, but higher than that generally
reported at these luminosities ($0.1-0.3$). 
The difference is most likely due to the 
lack of bias against obscured sources in the low-frequency-radio-selected 
3CRR sample. 

\item The multi-wavelength properties of many of the 3CRR NLRGs reveal
significantly ($10-1000 \times$) higher levels of intrinsic absorption 
(\nh(int)) than indicated by the X-ray hardness ratios. In such cases the use
of X-ray hardness ratio to correct for \nh(int) results in \lx~ values 
$\sim 10-100 \times$ too low.

\item We conclude that 8 of the NLRGs (50$\pm$18\%) are Compton thick 
(3C~13/68.2/266/324/356/368/437 and 4C~13.66). The high fraction
of CT sources (21\%) compared with visible and X-ray-selected samples
is a result of 
the lack of bias against heavily-obscured sources via radio-selection.
The ratio of unobscured to Compton-thin
($10^{22}<$ \nh(int) $< 1.5 \times 10^{24}$ cm$^{-2}$)
to CT (\nh(int)$> 1.5 \times 10^{24}$ cm$^{-2}$) is 2.5:1.4:1.

\item 
Assuming a random distribution of orientation
and a simple geometry in which the obscuring material is concentrated 
perpendicular to the radio axis, we deduce that obscuration in
the nuclear regions of high-z, radio-loud AGN includes:
a CT obscuring disk/torus extending $\sim~12$\deg~from the midplane,
additional obscuring material extending for another $\sim~18$\deg~with the
density decreasing away from the mid-plane, and the remaining
$\sim~60$\deg~is largely unobscured.
This last is consistent with previous estimates of torus/disk opening 
angles for high-luminosity AGN.

\item \lxlr~and L$[$OIII]/\lx,
in comparison with typical values for broad-line AGN,
provide a better measure of intrinsic absorption than the 
X-ray hardness ratio (HR).

\item The distribution of \nh(int) for the high-redshift 3CRR sample
peaks at \nh(int)$>10^{24}$ cm$^{-2}$, similar to the
results for lower-redshift 3CRR sources.

\item Given the edge-on nature of the host galaxies of at least 5
of the CT NLRGs (3C~13/266/324/368 and 4C~13.66), 
it is likely that host galaxy absorption  contributes
significantly to dust absorption signatures such as \tauSi.

\item The Compact Steep Spectrum (CSS) RLAGN 
(3C~43/186/190/241/287/318/454.0 and 4C~13.66) have a factor 
of $\sim 2.5$ lower \lxlr~and softer X-ray spectra ($\Gamma \sim 1.8-2$ 
cf. 1.7) than the non-CSS RLAGN in this complete sample. 

\end{enumerate}

\section*{Acknowledgements}
Support for this work was provided by the National Aeronautics and
Space Administration through \chandra\ Award Number G08-9106X, by the
\chandra\ X-ray Center, which is operated by the Smithsonian
Astrophysical Observatory for and on behalf of the National
Aeronautics Space Administration under contract NAS8-03060 
(\chandra\ X-ray Center) and by the Smithsonian Institution Endowment,
Scholarly Studies Program, fund \# 40488100HH0017.
The scientific results in this article are based to a significant degree 
on observations made by the \chandra~X-ray Observatory (\facility{CXO}).
We would like to thank Mark Avara and Margaret Yellen for
their early work on this project.

\bibliography{refs}
\bibliographystyle{apj}

\newpage
\begin{sidewaystable}[t]
\tablenum{1}
\caption{X-ray Observations and other Properties of the 3CRR High Redshift Sample }
\label{tb:obs}
\begin{minipage}{\textwidth}
\tiny{
\begin{tabular*}{0.9\textwidth}{@{\extracolsep{\fill}}llcllclcclllcc}
\hline\hline
Name & OBSID & Date Obs. &RA & Dec. & z & Exp. time & Source  & Galactic $N_H$ & Ref. & F(5 GHz) & log \rcd  &
log $\nu$\lr & Ref.\\
&  & UT & J2000.0 & J2000 & & ks &Type& $10^{20}\rm cm^{-2}$& X-ray & Jy (tot) && erg s$^{-1}$ (tot)  & Radio\\
\hline 
3CRR 009& \dataset[ADS/Sa.CXO\#Obs/1595]{1595}&		2001-06-10&	00:20:25.2&	+15:40:55&	2.009&	19.88&	QSO&	4.16&	1 &
0.546 & -2.04 & 44.91 &  14 \\
3CRR 013&	\dataset[ADS/Sa.CXO\#Obs/9241]{9241}	&	2008-06-01&	00:34:14.5&	+39:24:17&	1.351&	19.53&	NLRG&	6.39&	& 0.397
& -3.04 & 44.35 & 15 \\
3CRR 014&	\dataset[ADS/Sa.CXO\#Obs/9242]{9242}	&	2008-05-29&	00:36:06.5&	+18:37:59&	1.469&	3.00&		QSO&	4.12&
& 0.606 & -1.75 & 44.62 & 16 \\
3CRR 043&	\dataset[ADS/Sa.CXO\#Obs/9324]{9324}	&	2008-06-17&	01:29:59.8&	+23:38:20&	1.459&	3.04&		QSO/CSS&
7.13&&	1.082 & $<$-1.22 & 44.87 & 17 \\
3CRR 065&	\dataset[ADS/Sa.CXO\#Obs/9243]{9243}	&	2008-06-30&	02:23:43.2&	+40:00:52&	1.176&	20.91&	NLRG&	6.12&	& 0.765
& -3.17 & 44.48 & 15 \\
3CRR 068.1&	\dataset[ADS/Sa.CXO\#Obs/9244]{9244}	&	2008-02-10&	02:32:28.9&	+34:23:47&	1.238&	3.05&		QSO&	6.02&
& 0.824 & -2.87 & 44.57 & 18\\
3CRR 068.2&	\dataset[ADS/Sa.CXO\#Obs/9245]{9245}	&	2008-03-06&	02:34:23.8&	+31:34:17&	1.575&	19.88&	NLRG&	7.78&	& 0.179
& -2.63 & 44.17 & 15 \\
3CRR 181&	\dataset[ADS/Sa.CXO\#Obs/9246]{9246}	&	2009-02-12&	07:28:10.3&	+14:37:36&	1.382&	3.02	&	QSO&	6.83& &
0.655 & -2.03 & 44.59 & 19\\
3CRR 186&	\dataset[ADS/Sa.CXO\#Obs/3098]{3098}	&	2002-05-16&	07:44:17.4&	+37:53:17&	1.067&	34.44	&	QSO/CSS&	5.64&
2 & 0.377 & -1.38 & 44.07 & 17 \\
3CRR 190&	\dataset[ADS/Sa.CXO\#Obs/9247]{9247}	&	2007-12-31&	08:01:33.5&	+14:14:42&	1.195&	3.06	&	QSO/CSS&
2.65&	& 0.814 & -1.01 & 44.53 & 17 \\
3CRR 191&	\dataset[ADS/Sa.CXO\#Obs/5626]{5626}	&	2004-12-12&	08:04:47.9&     +10:15:23&	1.956&	19.77	&	QSO&	2.44&  3
& 0.457 & -0.99 & 44.81 & 20 \\
3CRR 204&	\dataset[ADS/Sa.CXO\#Obs/9248]{9248}	&	2008-01-13&	08:37:44.9&	+65:13:35&	1.112&	3.05	&	QSO&	4.27&
& 0.338 & -1.06 & 44.07 & 14 \\
3CRR 205&	\dataset[ADS/Sa.CXO\#Obs/9249]{9249}	&	2008-01-26&	08:39:06.4&	+57:54:17&	1.534&	96.72	&	QSO&	4.51&
5 & 0.665 & -1.51 & 44.71 & 21 \\
3CRR 208&	\dataset[ADS/Sa.CXO\#Obs/9250]{9250}	&	2008-01-08&	08:53:08.8&	+13:52:55&	1.110&	3.01	&	QSO&	3.59&
& 0.536 & -0.98 & 44.26 & 14 \\
3CRR 212&	\dataset[ADS/Sa.CXO\#Obs/434]{434}	&	2000-10-26&	08:58:41.5&	+14:09:44&	1.048&	18.05	&	QSO&	3.63&
4 & 0.884 & -0.69 & 44.42 & 22 \\
3CRR 239&	0306370701	& 2005-04-24	&	10:11:45.4&	+46:28:20&	1.781&		14 &	NLRG&
0.90&	5 & 0.328 & -2.82 & 44.56 & 15	\\
3CRR 241&	\dataset[ADS/Sa.CXO\#Obs/9251]{9251}	&	2008-03-13&	10:21:54.5&	+21:59:30&	1.617&	18.93	&	NLRG/CSS&	2.02&
5 & 0.338 & -2.05 & 44.47 & 23	\\
3CRR 245&	\dataset[ADS/Sa.CXO\#Obs/2136]{2136}	&	2001-02-12&	10:42:44.6&	+12:03:31&	1.029&	10.40	&	QSO&	2.87&
3 & 1.38 & +0.29 & 44.59 & 16 \\
3CRR 252&	\dataset[ADS/Sa.CXO\#Obs/9252]{9252}	&	2008-03-11&	11:11:33.0&	+35:40:42&	1.100&	19.45	&	NLRG&	1.73&
& 0.318 & -2.46 & 44.03 & 18\\
3CRR 266&	\dataset[ADS/Sa.CXO\#Obs/9253]{9253}	&	2008-02-16&	11:45:43.4&	+49:46:08&	1.275&	18.23	&	NLRG&	1.80&
& 0.318 & $<$-3.27 & 44.19 & 15 \\
3CRR 267&	\dataset[ADS/Sa.CXO\#Obs/9254]{9254}	&	2008-07-07&	11:49:56.5&	+12:47:19&	1.140&	19.18	&	NLRG&	2.90&
& 0.586 & -2.29 & 44.33 & 15\\
3CRR 268.4&	\dataset[ADS/Sa.CXO\#Obs/9325]{9325}	&	2009-02-23&	12:09:13.6&	+43:39:21&	1.398&	3.02	&	QSO&	1.30&
& 0.596 & -1.04 & 44.56 & 24 \\
3CRR 270.1&	\dataset[ADS/Sa.CXO\#Obs/9255]{9255}	&	2008-02-16&	12:20:33.9&	+33:43:12&	1.532&	9.67	&	QSO&	1.29&
6 & 0.864 & -0.55 & 44.82 & 16 \\
3CRR 287&	\dataset[ADS/Sa.CXO\#Obs/3103]{3103}	&	2002-01-06&	13:30:37.7&	+25:09:11&	1.055&	39.93	&	QSO/CSS&
1.08&	2 & 3.237 & $-$ & 44.99 & 25 \\
3CRR 294&	\dataset[ADS/Sa.CXO\#Obs/3207]{3207}	&	2002-02-27 &	14:06:44.0&	+34:11:25&	1.779&	123.63	&	NLRG&	1.21&
7 & 0.278 & -2.72 & 44.49 & 24 \\
3CRR 318&	\dataset[ADS/Sa.CXO\#Obs/9256]{9256}&		2008-04-05&	15:20:5.4&	+20:16:06&	1.574&	9.78	&	QSO/CSS&
4.01&		5 & 0.745 & $<$-0.86 & 44.79 & 26 \\
3CRR 322&	0028540301	& 2002-05-17	&	15:35:01.2&	+55:36:53&	1.681&	10.0/6.5$^{12}$&NLRG&
1.34&  8 & 0.457 & -3.18 & 44.64 & 27 \\
3CRR 324&	\dataset[ADS/Sa.CXO\#Obs/326]{326}&		2000-06-25&	15:49:48.9&	+21:25:38&	1.206&42.15	&	NLRG&	4.31& 9	
& 0.606 & $<$-3.64 & 44.41 & 28 \\
3CRR 325&	\dataset[ADS/Sa.CXO\#Obs/4818]{4818}&		2005-04-17&	15:49:58.4&     +62:41:22&	1.135&	28.66	&	QSO&	1.65& 10
& 0.824 & -2.53 & 44.48 & 18 \\
3CRR 356&	\dataset[ADS/Sa.CXO\#Obs/9257]{9257}&		2008-01-20&	17:24:19.0&	+50:57:40&	1.079&	19.87	&	NLRG&	2.76&
& 0.377 & -2.53 & 44.08 & 29 \\
4C 16.49&	\dataset[ADS/Sa.CXO\#Obs/9262]{9262}&		2008-01-21&	17:34:42.6&	+16:00:31&	1.880&	3.0	&	QSO&	6.64&
& 0.320 & -1.28 & 44.61 & 32\\
4C 13.66&	\dataset[ADS/Sa.CXO\#Obs/9263]{9263}&		2008-02-05&	18:01:38.9&	+13:51:24&	1.450&	19.90	&	NLRG/CSS&	11.15&
& 0.340 & $<$-2.23 & 44.36 & 31 \\
3CRR 368&	\dataset[ADS/Sa.CXO\#Obs/9258]{9258}&		2008-06-01&	18:05:6.3&	+11:01:33&	1.131&	19.90	&	NLRG&	9.03& &
0.209 & $<$-3.00 & 43.88 & 28 \\
3CRR 432&	\dataset[ADS/Sa.CXO\#Obs/5624]{5624}&		2005-01-07&	21:22:46.2&	+17:04:38&	1.785&	19.78	&	QSO&	7.34&
5,11 & 0.308 & -1.60 & 44.54 & 14 \\
3CRR 437&	\dataset[ADS/Sa.CXO\#Obs/9259]{9259}&		2008-01-07&	21:47:25.1&	+15:20:37&	1.480&	19.88	&	NLRG&	7.16&
& 0.874 & $<$-3.86 & 44.79 & 15 \\
3CRR 454.0	& 0306370201 &	2005-05-25&	22:51:34.7&	+18:48:40&	1.757&	16	&	QSO/CSS&	5.90&
5 & 0.784 & $<$-0.47 & 44.93 & 17\\
3CRR 469.1&	\dataset[ADS/Sa.CXO\#Obs/9260]{9260}&		2009-05-18&	23:55:23.3&	+79:55:20&	1.336&	20.18	&	NLRG&	13.74&
13 & 0.407 & -2.19 & 44.35 & 30 \\
3CRR 470&	\dataset[ADS/Sa.CXO\#Obs/9261]{9261}&		2008-03-03&	23:58:35.3&	+44:04:39&	1.653&	19.91	&	NLRG&	9.46&
& 0.546 & -2.43 & 44.70 & 15 \\
\hline\hline\\
\end{tabular*}
}\\
Notes:
1 \citet{2003MNRAS.338L...7F},
2 \citet{2008ApJ...684..811S},
3 \citet{2003A&A...401..505G},
4 \citet{2003ApJ...597..751A},
5 \citet{2008A&A...478..121S} (\xmm\ data),
6 \citet{2012ApJ...745...84W},
7 \citet{2001MNRAS.322L..11F,2003MNRAS.341..729F},
8 \citet{2004MNRAS.352..924B} (\xmm\ data, extended emission only),
9 \citet{2004ApJ...612..729H}, 
10 \cite{2009MNRAS.396.1929H} (used earlier, incorrect redshift of 0.86),
11 \citet{2006MNRAS.371...29E},
12 \xmm\ exposures in MOS/pn, after screening for periods of high background
13 \citet{2010MNRAS.401.1500L} (\xmm\ data),
14 \citet{1994AJ....108..766B},
15 \citet{1997MNRAS.292..758B},
16 \citet{1994A&AS..105..247A},
17 \citet{1998MNRAS.299..467L},
18 \citet{1997AJ....114.2292F},
19 \citet{1992MNRAS.257..353M},
20 \citet{1995A&AS..112..235A},
21 \citet{1984A&A...135...45L},
22 \citet{1991MNRAS.250..215A},
23 \citet{1985A&A...143..292F},
24 \citet{1992MNRAS.257..545L},
25 \citet{1989A&A...217...44F},
26 \citet{1991MNRAS.250..225S},
27 \citet{1995MNRAS.274..939L},
28 \citet{1998MNRAS.299..357B},
29 \citet{1993AJ....105.1690F},
30 \citet{1975MNRAS.173..309L},
31 \citet{1996MNRAS.279L..13R},
32 \citet{1993ApJS...87...63L}\\
\end{minipage}
\end{sidewaystable}

\newpage

\tablenum{2}
\begin{deluxetable}{lcrcllccccc}
\tabletypesize{\scriptsize}
\tablenum{2}
\tablewidth{0pt}
\tablecaption{X-ray Source Parameters$^1$ }
\tablehead{
  \colhead{Name} & \colhead{Type$^2$}& \colhead{Net Cts} &
  \colhead{Bkgrd. Cts} &  $\chi^2$& \colhead{$\Gamma$} & \colhead{$N_{H}(int)$} & \colhead{f(1 keV)$^4$} &
  \colhead{F(0.3-8 keV)$^4$} &
  \colhead{log L(0.3-8 keV)$^5$} & \colhead{HR$^6$} \\
  \colhead{} & \colhead{}  & \colhead{(0.3-8 keV)} &
  \colhead{(0.3-8 keV)} & \colhead{} &\colhead{} & \colhead{$10^{22}$cm$^{-2}$} &
\colhead{$10^{-6}$} &
  \colhead{$10^{-14}$} & \colhead{erg\,s$^{-1}$} & \colhead{$\frac{(H-S)}{(H+S)}$} 
}
\startdata
3C 009&  Q&   805.8$\pm$28.4& 1.19$\pm$0.08& 0.9 & 1.9& $<0.45$
&$  47.5^{+2.9}_{-2.8}$&  26.2$\pm$0.9& $45.85^{+0.01}_{-0.02}$& $-0.53^{+0.1}_{-0.1}$   \\ 
&&&& 0.8& 1.74$\pm$0.08 & $<0.28$ & 44.9$^{+2.2}_{-2.5}$ & 27.4$\pm$1.6 & \\
3C 013&  N/CT&    15.3$\pm$ 4.0& 0.68$\pm$0.06& 0.3 & 1.9 &$-$ &$   0.8_{ -0.5}^{+ 0.5}$&   0.4$\pm$0.2& $43.6_{-0.4}^{+0.2}$& $
-0.51^{+0.2}_{-0.2}$       \\ 
3C 014&  Q&   238.9$\pm$15.5& 0.14$\pm$0.03&0.6&1.9&  $0.7_{-0.3}^{+ 0.4}$&$ 131.8_{-14.9}^{+16.0}$&  73.0$\pm$3.9&
$45.97_{-0.02}^{+0.03}$ & $-0.38^{+0.1}_{-0.1}$        \\ 
3C 043&  Q/C&   170.8$\pm$13.1& 0.16$\pm$0.03&0.5&1.9&  $<1.5$&$  86.6_{-12.0}^{+13.4}$&  47.7$\pm$3.2& $45.77_{-0.03}^{+0.03}$& $
-0.40^{+0.1}_{-0.1}$       \\
3C 065&  N&   205.2$\pm$14.4& 0.81$\pm$0.06&0.6 &1.9& $9.3_{-1.8}^{+ 2.2}$&$  45.4_{ -7.1}^{+ 7.6}$&  25.1$\pm$0.3& $45.26_{-0.01}^{+0.01}$& $ 
+0.23^{+0.1}_{-0.1}$        \\ 
3C 068.1&Q&    43.7$\pm$ 6.6& 0.29$\pm$0.04&0.4 &1.9& $9.0_{-4.6}^{+ 7.3}$&$  57.4_{-23.6}^{+28.9}$&  131.8$\pm$1.6& $45.42_{-0.02}^{+0.02}$& $ 
+0.13^{+0.1}_{-0.2}$        \\
3C 068.2&N/CT&  8.2$\pm$ 3.0& 0.77$\pm$0.06&0.2&1.9&  $- $&$   0.37_{-0.35}^{+0.35}$&   0.21$\pm$0.16& $43.5_{-0.6}^{+0.2}
$& $+0.33^{+0.4}_{-0.3}$           \\ 
3C 181&  Q&   188.8$\pm$13.8& 0.16$\pm$0.03&0.7&1.9&  $<0.6$& $ 86.1_{-8.3}^{+10.4}$&  47.3$\pm$4.2&
$45.71_{-0.04}^{+0.04}$& $-0.53^{+0.1}_{-0.1}$ \\
3C 186$^7$ & Q/C&  1984.4$\pm$44.6& 4.65$\pm$0.15&1.3&1.9&  $<0.01 $& $ 70.5_{-3.2}^{+3.5}$&  39.0$\pm$1.0&
$45.35_{-0.01}^{+0.01}$& $-0.63^{+0.02}_{-0.02}$ \\ 
&&&&1.2&$2.1\pm0.1$ & $<0.03$ &$ 72.9_{-1.9}^{+3.6}$ & $36.8\pm1.0$ &&\\
3C 190&  Q/C&   172.8$\pm$13.2& 0.16$\pm$0.03&0.9&1.9&  $0.4_{-0.2}^{+ 0.3}$&$  80.5_{-11.0}^{+11.7}$&  45.2$\pm$3.6&
$45.54_{-0.04}^{+0.03}$& $-0.52^{+0.1}_{-0.1}$  \\
3C 191&  Q&   824.2$\pm$28.7& 0.76$\pm$0.06&0.8&1.9& $<0.45$ 
&$  51.4_{-3.0}^{+3.1}$&  28.4$\pm$1.1& $45.86_{-0.02}^{+0.01}$& $-0.52^{+0.03}_{-0.03}$        \\
&&&&0.7&1.70$\pm$0.08 & $<0.25$ & $47.5^{+3.1}_{-2.1}$ & 29.6$\pm$1.9 && \\ 
3C 204&  Q&   358.8$\pm$19.0& 0.20$\pm$0.03&0.6&1.9&  $<0.4$ & $ 168.9_{-14.0}^{+14.4}$&  92.6$\pm$5.2&
$45.86_{-0.02}^{+0.01}$& $-0.57^{+0.1}_{-0.1}$       \\ 
3C 205$^8$&  Q&  1006.5$\pm$31.7& 0.51$\pm$0.05&1.0&1.9&  $0.42_{-0.14}^{+0.16}$ & $164.4_{-8.6}^{+8.8}$&  90.8$\pm$2.5&
$46.11_{-0.02}^{+0.01}$& $-0.43^{+0.1}_{-0.1}$ \\
&&&&0.6 & $1.60^{+0.08}_{-0.05}$ & $<0.4$ & $134.5^{+10.0}_{-5.3}$ & 91.4$\pm$4.8 && \\
3C 208&  Q&   280.8$\pm$16.8& 0.20$\pm$0.03&0.8&1.9&  $<0.5 $ & $ 126.6_{-10.0}^{+14.9}$ & 69.6$\pm$5.0&
$45.65_{-0.04}^{+0.03}$& $-0.47^{+0.1}_{-0.1}$ \\
3C 212&  Q& 3944.1$\pm$62.8& 0.92$\pm$0.07&0.8&1.9&  $0.46_{-0.03}^{+ 0.03}$& $ 330.2_{-8.1}^{+8.2}$& 182.5$\pm$1.9&
$46.00_{-0.01}^{+0.01}$& $-0.49^{+0.01}_{-0.01}$ \\ 
&&&&0.6&1.68$\pm$0.04 & $0.32^{+0.04}_{-0.03}$ &  $290.8^{+10.7}_{-10.3}$ &183.5$\pm$3.4 && \\ 
3C 239$^9$&  N &      $-$        & $-$& $-$& $-$& $-$  & $-$ & $-$ & $44.69_{-0.06}^{+0.06}$ & $-0.7$  \\   
3C 241&  N/C&   147.2$\pm$12.2& 0.76$\pm$0.06&0.7&1.9&  $6.2_{-1.7}^{+ 2.4}$&$  22.7_{-4.0}^{+4.6}$&   12.6$\pm$0.4&
$45.30_{-0.01}^{+0.01}$& $-0.05^{+0.1}_{-0.1}$ \\
3C 245&Q&  2067.4$\pm$45.5& 0.65$\pm$0.07&1.4&1.4&  $<0.09$  &$ 236.0_{-8.4}^{+8.6}$& 130.3$\pm$2.6&
$45.84_{-0.01}^{+0.01}$& $-0.57^{+0.1}_{-0.1}$ \\ 
&&&& 1.0 & $1.65^{+0.05}_{-0.04}$ & $<0.02$& $218.6^{+8.2}_{-4.4}$ & 141.4$\pm$5.8 && \\
3C 252&  N&    89.6$\pm$ 9.5& 1.45$\pm$0.09&1.1&1.9& $10.5_{-3.9}^{+ 7.5}$&$  21.5_{ -6.4}^{+ 9.8}$&  11.9$\pm$0.2&
$44.87_{-0.01}^{+0.01}$& $ +0.43^{+0.1}_{-0.1}$  \\ 
3C 266&  N/CT&    19.2$\pm$ 4.5& 0.77$\pm$0.06&0.3&1.9& $-$ &   $ 0.71_{ -0.44}^{+ 0.44}$&   0.36$\pm$0.23& $43.51_{-0.44}^{+0.21}$& 
$+0.19^{+0.2}_{-0.2}$      \\ 
3C 267&  N&   167.2$\pm$13.0& 0.82$\pm$0.06&1.3&1.9& $10.8_{-2.6}^{+ 3.7}$&$  39.8_{ -7.8}^{+ 9.5}$&  22.0$\pm$0.4&
$45.18_{-0.01}^{+0.01}$& $+0.31^{+0.1}_{-0.1}$ \\ 
3C 268.4&Q&   291.8$\pm$17.1& 0.18$\pm$0.03&1.2&1.9&  $0.4_{-0.2}^{+0.3}$ & $ 142.7_{-14.7}^{+15.8}$&  78.6$\pm$5.0&
$45.94_{-0.02}^{+0.03}$& $-0.40^{+0.1}_{-0.1}$ \\ 
3C 270.1&Q&   734.6$\pm$27.1& 0.45$\pm$0.05&1.2&1.9&  $<0.36$ & $ 95.5_{-4.6}^{+4.6}$&  53.2$\pm$2.5&
$45.87_{-0.02}^{+0.02}$& $-0.53^{+0.1}_{-0.1}$        \\ 
&&&& 1.1 &$1.69\pm0.08$ & $<0.24$ & $91.2^{+6.1}_{-4.3}$ & 57.6$\pm$4.3&&\\
3C 287&Q/C&  3862.0$\pm$62.2& 2.01$\pm$0.14&0.6&1.9&  $<0.06$ & $ 120.0_{-2.9}^{+2.9}$&  66.3$\pm$1.2&
$45.57_{-0.01}^{+0.01}$& $-0.63^{+0.01}_{-0.01}$ \\ 
&&&&0.6&$1.82^{-0.04}_{+0.05}$ &$<0.09$ & $115.3^{+4.0}_{-3.2}$ & 66.6$\pm$1.7 && \\ 
3C 294$^{10}$ & N/CT &    202.9$\pm$14.4& 3.08$\pm$0.13&3.6&1.9&  $-$&$   0.61$&   0.34$\pm$0.06& $43.83_{-0.08}^{+0.08}$&
$+0.52^{+0.06}_{-0.06}$        \\ 
3C 318&  Q/C&   267.5$\pm$16.4& 0.50$\pm$0.05&0.5&1.9&  $<0.8 $ & $ 38.8_{-3.5}^{+4.2}$&  21.4$\pm$1.6&
$45.50_{-0.03}^{+0.04}$& $-0.50^{+0.1}_{-0.1}$  \\
3C 322$^{11}$&  N&   17.6$\pm$6.5   & 12.4$\pm$3.5 & $-$ & $-$ & $-$ & $-$& $<1.5$ & $<44.44$ & $-$  \\    
3C 324$^{12}$&  N/CT&    45.2$\pm$ 6.9& 2.80$\pm$0.12&0.4&1.9&  $-$ & $ 0.9_{-0.2}^{+0.2}$&   0.49$\pm$0.11&
$43.58_{-0.11}^{+0.09}$& $-0.28^{+0.16}_{-0.13}$ \\
3C 325&  Q&   360.1$\pm$19.0& 0.92$\pm$0.07&0.8&1.9&  $6.2_{-0.9}^{+ 1.0}$&$ 44.7_{-4.9}^{+5.2}$&  24.6$\pm$0.3&
$45.22_{-0.01}^{+0.01}$& $+0.05^{+0.06}_{-0.05}$        \\ 
3C 356&  N/CT&    26.1$\pm$ 5.2& 0.87$\pm$0.07&0.5&1.9&  $-$ &$   0.9_{-0.4}^{+0.4}$&   0.52$\pm$0.24& $43.49_{-0.27}^{+0.17}$&
 $+0.33^{+0.2}_{-0.2}$     \\ 
4C 16.49&Q&  183.8$\pm$13.6& 0.22$\pm$0.03&0.7 &1.9 & $<0.5 $&$ 83.7_{-6.8}^{+11.7}$&  46.9$\pm$4.0&
$46.03_{-0.04}^{+0.04}$& $-0.54^{+0.1}_{-0.1}$    \\  
4C 13.66&N/C/CT&    20.0$\pm$ 4.6& 0.98$\pm$0.07&0.7&1.9&  $-$  & $ 1.2_{-0.6}^{+0.6}$&   0.61$\pm$0.23& $43.87_{-0.20}^{+0.14}$& 
$-0.54^{+0.2}_{-0.2}$      \\ 
3C 368$^{12}$&  N/CT&    17.1$\pm$ 4.2& 0.86$\pm$0.07&0.1&1.9&  $-$   &$  1.0_{-0.6}^{+ 0.6}$&  0.53$\pm$0.24& $43.55_{-0.26}^{+0.16}$& 
$-0.34^{+0.2}_{-0.3}$      \\ 
3C 432&  Q&   771.2$\pm$27.8& 0.76$\pm$0.06&0.8&1.9&  $<0.6$&$  55.6_{-3.4}^{+3.5}$&  30.7$\pm$0.9&
$45.79_{-0.01}^{+0.02}$& $-0.50^{+0.03}_{-0.03}$ \\ 
&&&&0.8& $1.74^{+0.11}_{-0.07}$& $<0.67$ & $50.3^{+4.9}_{-2.5}$ & 30.6$\pm$1.7 & & \\
3C 437&  N/CT& 9.8$\pm$ 3.3& 0.84$\pm$0.06&0.4&1.9&   $-$ & $ 0.38_{ -0.35}^{+ 0.35}$&   0.19$\pm$0.17&
$43.39_{-0.98}^{+0.28}$& $+0.28^{+0.3}_{-0.3}$ \\ \hline\\  
3C 454.0$^9$&Q/C& $-$  & $-$ & $-$& $-$&$<$0.13& $-$ & $-$ & $45.70_{-0.03}^{+0.03}$ & $-0.3$  \\
3C 469.1&N&    80.9$\pm$ 9.1& 1.12$\pm$0.08&0.3&1.9& $26.9_{-8.9}^{+14.8}$&$  32.6_{-10.1}^{+14.3}$&   18.0$\pm$0.2&
$45.26_{-0.01}^{+0.01}$& $+0.61^{+0.1}_{-0.1}$ \\ 
3C 470&  N&    55.2$\pm$ 7.5& 0.85$\pm$0.07&0.2&1.9&$50.7_{-15.5}^{+22.8}$&$  29.1_{-10.4}^{+13.2}$&   16.0$\pm$0.2&
$45.43_{-0.01}^{+0.01}$& $+0.74^{+0.1}_{-0.1}$ \\ 
\enddata
\label{tb:flux}
\tablecomments{
1) X-ray data were fit with a power law ($\Gamma=1.9$) + Galactic
  absorption, plus intrinsic equivalent hydrogen column density 
($N_{H}(int)$).
Generally no constraints could be placed on $N_H$(int) when
the net counts were $<50$.\\
2) Source Classification: Q = quasar; N = Narrow Line Radio Galaxy (NLRG); C = 
Compact Steep Spectrum (CSS) source; CT = Compton Thick candidate  \\
4) Flux densities (in units of pht\,cm$^{-2}$\,s$^{-1}\,$keV$^{-1}$) and fluxes (in units of erg\,cm$^{-2}$\,s$^{-1}$)
are quoted with 1$\sigma$ errors and based on the best 
fit spectral model.   Fits to all NLRGs with \gax 50cts required intrinsic ($N_{H}(int)$)
absorption. Upper limits to \nh(int) are quoted at 3$\sigma$.
\\
5) L(0.3-8keV) is determined in the rest frame including a correction
for any significant \nh(int) \\
6) Hardness ratios are calculated using BEHR \citep{2006ApJ...652..610P} \\
7) Detailed spectral fits for this source and its surrounding cluster
are presented in \citet{2005ApJ...632..110S,2010ApJ...722..102S} \\
8) The X-ray spectrum is flat, $\Gamma \sim 1.6$, 
this results in an apparent \nh(int) for the $\Gamma = 1.9$ fit.\\
9) Equivalent \chandra\ luminosity and HR were 
derived  based on published \xmm\ spectral 
fits which reported no counts or flux 
\citep{2008A&A...478..121S,2010ApJ...722..102S}.\\
10) A single power law does not provide a good fit to the data.
A detailed spectral fit \citep{2003MNRAS.341..729F} 
yields \nh=$8.4^{+1.1}_{-0.9} \times 10^{23}$ cm$^{-2}$,
dominated by a reflection component. This source is not included
as a CT candidate in Section~\ref{sec:CT}.\\
11) The upper limit was determined directly from the 
\xmm~dataset using an on-source circle of radius 8''.
Due to the weak detection of extended emission aligned with the lobes, the
counts are treated as an upper limit to any true core emission. \\
12) Hardness ratios are significantly different using a smaller 
circle (1'' radius): 3C 324: $0.13^{+0.16}_{-0.20}$; 
3C 368: $-0.01^{+0.35}_{-0.25}$ 
}
\end{deluxetable}

\begin{deluxetable}{lc}
\tablenum{3}
\tablecaption{\nh(int) for low count sources
estimated from Figure~\ref{fg:LxvsHR}.}
\tabletypesize{\footnotesize}
\tablewidth{0pt}
\tablehead{
\vspace{0.2cm}\\
\colhead{  Name  } & \colhead{\nh(int)$^1$} \\
& \colhead{$10^{24}$ cm$^{-2}$} 
} 
\startdata
3C 13 & 1.8 \\
3C 68.2 & 2.0 \\
3C 266 & 1.8 \\
3C 294 & 1.4 \\
3C 324 & 1.8 \\
3C 356 & 1.9 \\
3C 368 & 1.9 \\
3C 437 & 2.0 \\
4C 13.66 & 1.4 \\
\enddata 
\label{tb:f4nh}
\tablecomments{1: Uncertainty is $\pm 0.2$, accounting for  
1$<$z$<$2, 1.5$<\Gamma<$2.2.}
\end{deluxetable}

\clearpage

\clearpage

\clearpage

\clearpage

\clearpage

\clearpage

\clearpage

\clearpage

\clearpage

\clearpage

\clearpage

\end{document}